\documentclass[letterpaper,draftclsnofoot,onecolumn,11pt]{IEEEtran}
\usepackage{amssymb, amsmath, epsfig, cite, enumitem, amsthm, bbm}
\usepackage{comment,color}
\newtheorem{mydef}{Definition}
\newtheorem{thm}{Theorem}
\newtheorem{lm}{Lemma}
\newtheorem{conj}{Conjecture}

\newcommand  {\ssw} {\mathrm{ssw}}
\newcommand  {\BSC} {\mathrm{BSC}}
\newcommand  {\BEC} {\mathrm{BEC}}
\newcommand  {\bc}  {{\boldsymbol c}}
\newcommand  {\bb}  {{\boldsymbol b}}
\newcommand  {\bw}  {{\boldsymbol w}}
\newcommand  {\bx}  {{\boldsymbol x}}
\newcommand  {\by}  {{\boldsymbol y}}
\newcommand  {\bu}  {{\boldsymbol u}}
\newcommand  {\bs}  {{\boldsymbol s}}

\newcommand  {\bQ}  {{\boldsymbol Q}}
\newcommand  {\bS}  {{\boldsymbol S}}
\newcommand  {\bX}  {{\boldsymbol X}}
\newcommand  {\bY}  {{\boldsymbol Y}}
\newcommand  {\mv} {{\mathbbm{v}}}
\newcommand  {\mA} {{\mathbbm{A}}}
\newcommand  {\mB} {{\mathbbm{B}}}
\newcommand  {\mI} {{\mathbbm{I}}}
\newcommand  {\mOne} {{\mathbbm{1}}}
\newcommand  {\cluster}  {{\mathrm{cluster}}}
\newcommand {\Rs} {{\mathbb{R}}}
\newcommand {\Ef} {{\mathfrak{E}}}
\newcommand {\setV} {{\mathcal{V}}}
\newcommand {\setE} {{\mathcal{E}}}
\newcommand {\setQ} {{\mathcal{Q}}}
\newcommand {\setG} {{\mathcal{G}}}
\newcommand {\setB} {{\mathcal{B}}}
\newcommand {\setO} {{\mathcal{O}}}
\newcommand {\setR} {{\mathcal{R}}}
\newcommand {\setW} {{\mathcal{W}}}
\newcommand {\setS} {{\mathcal{S}}}
\newcommand {\setX} {{\mathcal{X}}}
\newcommand {\setY} {{\mathcal{Y}}}

\begin{document}

\title{Skip-Sliding Window Codes}

\author{Ting-Yi~Wu,
	Anshoo Tandon,
 	Lav~R.~Varshney,~\IEEEmembership{Senior Member,~IEEE}, and
	Mehul Motani,~\IEEEmembership{Fellow,~IEEE}
\thanks{T.-Y.\ Wu and L.~R.\ Varshney are with the Coordinated Science Laboratory, University of Illinois at Urbana-Champaign, Urbana, IL 61801, USA 
(e-mail: \{tywu, varshney\}@illinois.edu).
A.~Tandon and M.~Motani are with the Department of Electrical and Computer Engineering, National University of Singapore, Singapore 117583
(e-mail: \{anshoo.tandon@gmail.com; motani@nus.edu.sg\}).
}
}

\maketitle
		 
\begin{abstract}
Constrained coding is used widely in digital communication and storage systems. In this paper, we study a generalized sliding window constraint called the skip-sliding window. A skip-sliding window (SSW) code is defined in terms of the length $L$ of a sliding window, skip length $J$, and cost constraint $E$ in each sliding window. Each valid codeword of length $L + kJ$ is determined by $k+1$ windows of length $L$ where window $i$ starts at $(iJ + 1)$th symbol for all non-negative integers $i$ such that $i \leq k$; and the cost constraint $E$ in each window must be satisfied. In this work, two methods are given to enumerate the size of SSW codes and further refinements are made to reduce the enumeration complexity. Using the proposed enumeration methods, the noiseless capacity of binary SSW codes is determined and observations such as greater capacity than other classes of codes are made. Moreover, some noisy capacity bounds are given. SSW coding constraints arise in various applications including simultaneous energy and information transfer. 
\end{abstract}

\section{Introduction}
Constrained coding losslessly maps a set of unconstrained sequences into a set of sequences 
that satisfy certain constraints, and has been extensively used in several applications. 
To alleviate timing errors due to the rapid change of stored bits in magnetic and optical storage,
binary {\it runlength-limited codes} \cite{Immink1990, Immink2004} are employed to insert a run of zeros between consecutive ones.
In simultaneous information and energy transmission \cite{UlukusYESZGH2015}, 
a minimal number of ones in subsequences of transmitted codewords is required so as
to carry enough energy while transmitting information \cite{RosnesBY2012, FouladgarSE2014, TandonMV2014a, TandonMV2016, TandonKM2017}.
Asynchronous communication necessitates codes with {\it heavy/light Hamming weight} \cite{CohenST2010, BachocCCST2011}.

\begin{figure}
    \centering
    \includegraphics[width=5in]{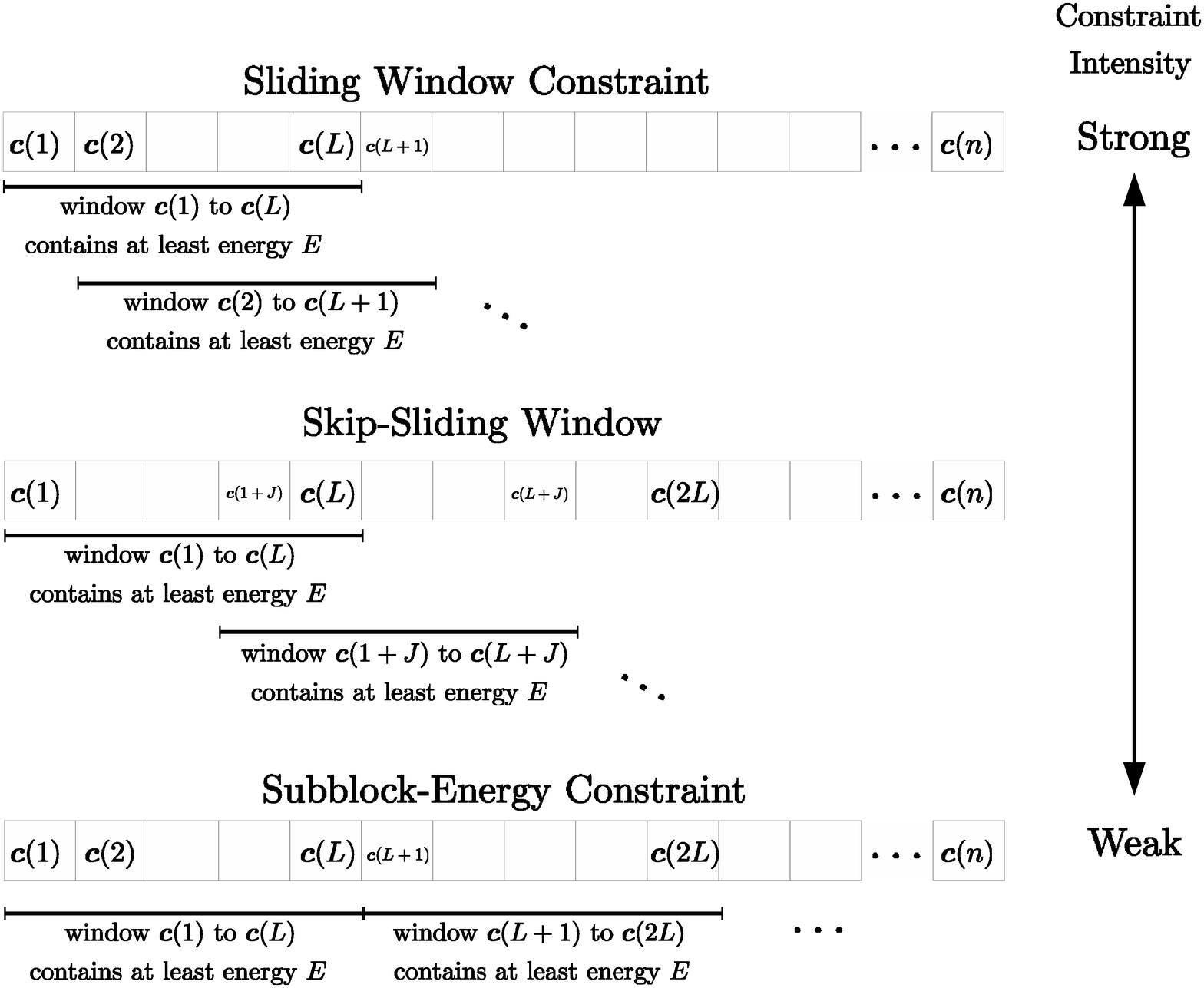}
    \caption{Sliding window constraint codes, skip-sliding window codes, and subblock-energy constraint codes.
    As the intensity of the coding constraint increase ($J$ decrease), 
    it is expected that the capacity (noiseless or noisy) of the code will decrease.
    However, the skip-sliding window codes may violate this intuition. 
    }
    \label{fig:excc}
\end{figure}

Two basic constrained coding strategies have been developed for simultaneous information and energy communication: 
{\it sliding window constraint} (SWC) codes \cite{RosnesBY2012, FouladgarSE2014, TandonMV2014a} 
and {\it subblock-energy constraint} (SEC) codes \cite{TandonMV2016, TandonKM2017}.
As Fig.~\ref{fig:excc} shows, SWC codes restrict the energy of every consecutive $L$ symbols to be no less than $E$ to 
guarantee enough energy is conveyed after transmitting any symbol. 
The sliding-window constraint enables SWC codes to convey energy to meet real-time delivery requirements, 
but also reduces the number of valid SWC codewords and therefore the information capacity. 
When there are energy buffers (batteries), energy transmission need not be so constrained at the level of individual transmitted symbols, 
and so SEC codes only restrict the energy of non-overlapping subblocks to be no less than $E$; 
this leads to more allowable codewords and capacity.  

This work introduces a new intermediate type of constrained code
that generalizes both SWC and SEC codes. 
Instead of assuring the energy constraint on consecutive $L$ symbols for all sliding windows, 
the proposed constrained code that we call {\it skip-sliding window} (SSW) codes 
loosen the SWC constraint by lifting the energy constraint for those sliding windows 
that do not start at $\left(iJ+1\right)$th symbols, where $J$ is a fixed integer and $i$ is any non-negative integer.  
It is immediate that an SSW code reduces to an SWC code when $J=1$, 
and to an SEC code when $J=L$.  
In this sense, SWC and SEC codes are two ends of a spectrum of SSW codes as Fig.~\ref{fig:excc} shows.

Note that although SSW codes are inspired by constrained codes for simultaneous information and energy transmission,
they may also be useful in several other areas where sliding window constraints arise.
Genome sequences may be assembled using de Bruijn graphs \cite{CompeauPT2011, Bruijn1946}, 
in which constrained overlapped genome subsequences form a graph that can be related to a sliding window constraint. 
Several task scheduling problems state that machines process jobs in a sliding-window manner when loading is limited \cite{BeldiceanuD2010}; 
sequencing of tasks to satisfy the loading constraint forms an SSW code.  Other topics, such as {\it Rauzy graphs} \cite[Sec.~3.4.2]{Salimov2010, Rigo2014} 
in formal languages and {\it regular constraints} \cite{HoevePRS2006} in constraint programming \cite{Apt2003} 
may also be treated from an SSW code perspective. 

The main contributions of this paper are as follows.  
\begin{enumerate}
\item We define SSW codes and characterize basic properties to build the mathematical foundation for SSW-related applications. 
\item We introduce two methods to enumerate valid codewords of binary SSW codes where the cost constraint $E$ is defined as the least Hamming weight $W$: one is based on the adjacency matrix of a modified de Bruijn graph \cite{Bruijn1946}
which enumerates the SSW code in complexity $O\Big(\big(\sum_{i=W}^L{L \choose i}\big)^2\Big)$, whereas the other uses the Goulden-Jackson cluster method \cite{GouldenJ1979} which enumerates the SSW code with $L=\ell J$ for some positive integer $\ell$ in complexity $O\Big(\big(\sum_{i=0}^{W-1}{L \choose i}\big)^2\Big)$. 
A modified Goulden-Jackson cluster method is further proposed which is proved to be equivalent to the first method when $L=\ell J$ for some positive integer $\ell$.  
\item When $L=\ell J$ for some positive integer $\ell$, refinements of both enumeration methods are given to lower the complexity. The refinement of the first enumeration method reduces its complexity to be $O\big(\min \{J+1, W+1\}^{\ell-1}\big)$, and the refinement of the second enumeration method reduces its complexity to be $O\big(\min \{J, W-1\}^{\ell-1}\big)$.
\item Properties of the noiseless capacity of SSW are proven, and some interesting and useful observations from numerical simulations are given. 
In particular, SSW codes can surprisingly achieve higher noiseless capacity than SEC codes do.
\item Several noisy capacity bounds over the binary symmetric channel (BSC) and the binary erasure channel (BEC) are given for comprehensiveness.
\end{enumerate}

The rest of this paper is organized as follows. 
Section \ref{sec:ssw} introduces SSW codes and their noiseless capacity. 
Enumeration methods are given in Section \ref{sec:enumeration} 
and refined enumeration methods are further derived in Section \ref{sec:refined}.
Section \ref{sec:properties} gives properties and numerical results on SSW codes in the noiseless case.
Some noisy capacity bounds and their numerical evaluations are given in Section \ref{sec:noisy}.
Section \ref{sec:con} summarizes and concludes.  

\section{Skip-Sliding Window Codes}\label{sec:ssw}
Let us consider $q$-ary sequences where each symbol in the sequence is drawn from $\setQ\triangleq\{0, 1,\ldots, q-1 \}$ and
define a cost function $\Ef(\cdot)$ which maps each symbol to a real value as 
$\Ef  :\setQ \rightarrow \Rs$.
A $q$-ary SSW sequence with a window length $L$, a skip length $J$, and a minimal cost $E$, 
denoted as an $(L, J, E)_q$-SSW sequence, 
guarantees the sum of the cost of the $L$ consecutive symbols which start at the $\left(iJ+1\right)$th symbol
to be no less than $E$ for all non-negative integers $i$.
\begin{mydef}
Given positive integers $L$ and $J$ such that $L\geq J$,
a cost function $\Ef  :\setQ \rightarrow \Rs$, and the minimal cost $E$,  
a $q$-ary sequence of length $n=L+kJ$ is said to be an \emph{$(L, J, E)_q$-SSW sequence} if 
\begin{equation} \label{eqn:sswq}
\sum_{i=1}^{L}\Ef \left(\bc(i+mJ)\right)\geq E \mbox{ for all } 0\leq m\leq k,
\end{equation}
where $\bc(i)$ denotes the $i$th symbol of the sequence $\bc$.
\end{mydef}
\begin{mydef}
The collection of all $(L, J, E)_q$-SSW sequences of length $n$ form the \emph{$(L, J, E)_q$-SSW code} of length $n$. 
\end{mydef}

Since binary sequences are of particular interest, we largely focus on $(L, J, E)_2$-SSW sequences in the sequel. 
For simplicity, the cost function is taken as the Hamming weight of the binary symbol, $\Ef (\bc(i))=\bc (i)$, 
and the cost constraint $E$ is replaced by $W$ to specifically denote the Hamming weight. 
\begin{mydef}
Given positive integers $L$, $J$, and $W$, 
such that $L \geq W$ and $L\geq J$,
a binary sequence $\bc$ of length $n=L+kJ$ is said to be an \emph{$(L, J, W)$-SSW sequence} if 
\begin{equation}\label{eqn:ssw}
\sum_{i=1}^{L}\bc(i+mJ)\geq W \mbox{ for all } 0\leq m \leq k,
\end{equation}
where $\bc(i)$ denotes the $i$th bit value of the binary sequence $\bc$.
\end{mydef}

Let $M_\ssw^{(L, J, W)}(L+kJ)$ denote the number of the $(L, J, W)$-SSW sequences of length $L+kJ$ for some non-negative integer $k$.
Our interest is in finding $M_\ssw^{(L, J, W)}(L+kJ)$, but 
especially the noiseless capacity of binary skip-sliding codes, 
\begin{equation}\label{eqn:capacity}
C_\ssw^{(L, J, W)}\triangleq \lim_{k\to\infty}\frac{\log M_\ssw^{(L, J, W)}(L+kJ)}{L+kJ}.
\end{equation}
The next two sections will introduce several ways to enumerate $M_\ssw^{(L, J, W)}(L+kJ)$.

Before closing this section, we present the following theorem which states that  
any $(L, J, W)$-SSW sequence with $L=\ell J$ for some $\ell>0$ is equivalent to a
$q$-ary $(\ell, 1, E)_q$-SSW sequence such that $q=2^J$. 
\begin{thm} \label{thm:qary2binary}
For any binary $(L, J, W)$-SSW code where $L=\ell J$ for some positive integer $\ell$, 
there is an equivalent $2^J$-ary $(\ell, 1, E)_{2^J}$-SSW code. 
\end{thm}
\begin{IEEEproof}
Let $w(\bc)$ be the Hamming weight of the binary string $\bc$ and $(i)_J$ be the binary representation of length $J$ for the non-negative integer 
$i$.\footnote{For example, $(5)_4=0101$, hence the $w\left((5)_4\right)=w(0101)=2$.}
Then we can construct a $2^J$-ary $(\ell, 1, E)_{2^J}$-SSW code such that $E=W$, $\setQ=\{0, 1, \ldots , 2^J-1\}$ and 
$\Ef (i)=w\left( (i)_J\right)$ for all $i\in \setQ$.
Hence, any $2^J$-ary $(\ell, 1, E)_{2^J}$-SSW sequence of length $n$
can be transformed to be a binary $(L, J, W)$-SSW sequence of length $nJ$ by
representing each symbol in binary, i.e.,  
$\bc \in (\ell , 1, E)_{2^J}$-code of length $n$ if and only if $(\bc(1))_J (\bc(2))_J\cdots (\bc(n))_J \in (L, J, W)$\mbox{-code of length $nJ$}.
\end{IEEEproof}

\section{Enumeration Methods}\label{sec:enumeration}
To enumerate $M_\ssw^{(L, J, W)}(n)$, we consider its generating function $g(x)$, such that
\begin{equation}\label{eqn:GF}
g(x)=\sum_{n=0}^{\infty}M_\ssw^{(L, J, W)}(n)x^n.
\end{equation}
Note that $M_\ssw^{(L, J, W)}(n)=0$ if $n\neq L+kJ$ for any non-negative integer $k$.

\subsection{Finite State Machine}\label{sec:FSM}
\begin{figure}
    \centering
    \includegraphics[width=2in]{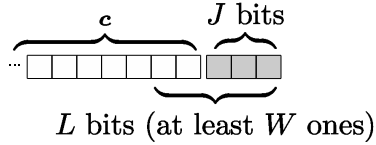}
    \caption{Extending an $(L, J, W)$-SSW sequence $\bc$.}
    \label{fig:exseq}
\end{figure}

To \emph{extend} an $(L, J, W)$-SSW sequence $\bc$, as Fig.~\ref{fig:exseq} shows,
the incoming $J$ bits and the last $L-J$ bits of $\bc$ must contain at least $W$ ones.
Hence, the incoming $J$ bits and the last $L-J$ bits of $\bc$ can determine if the extended sequence is a valid $(L, J, W)$-SSW sequence,
which indicates that the finite state machine (FSM) with $L$-bit states can represent all possible $(L, J, W)$-SSW sequences.

\begin{figure}
    \centering
    \includegraphics[width=3in]{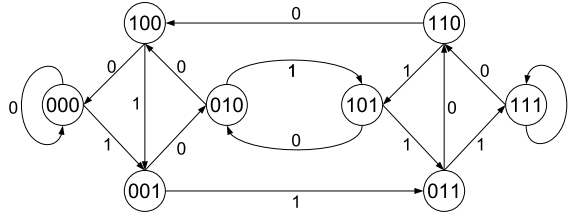}
    \caption{An example of de Bruijn graph of order $3$, which is also the FSM of the $(3, 1, 0)$-SSW code.}
    \label{fig:BruijnGraph}
\end{figure}

Let us consider a directed graph $G(\setV , \setE )$ with vertex set $\setV$ and directed edge set $\setE$,
which contains all $L$-bit vertices, i.e.,
\begin{equation*}
\setV=\{[b_1\cdots b_L]: b_i\in\{0,1\} \mbox{ for all } 1\leq i\leq L \}, 
\end{equation*}
and the vertex $[b_1\cdots b_L]$ can transit to the vertex  $[b'_1\cdots b'_L]$ if 
$b_{i+1}=b'_{i}$ for all $1\leq i \leq L-1$, i.e.,
\begin{equation*}
([a \,b_1\cdots b_{L-1}], [b_1\cdots b_{L-1}a'])\in \setE
\end{equation*}
for all $a$, $a'$, and $b_i\in\{0,1\}$.
Such a graph $G(\setV , \setE )$ is called the de Bruijn graph of order $L$ \cite{Bruijn1946}. 
An example of a de Bruijn graph of order $3$ is depicted as Fig.~\ref{fig:BruijnGraph}.
Since the states in the de Bruijn graph of order $L$ represent the latest $L$ bits of the incoming path, 
the de Bruijn graph of order $L$ can be treated as an FSM of an $(L, 1, 0)$-SSW code.
Hence Fig.~\ref{fig:BruijnGraph} is also the FSM of an $(3, 1, 0)$-SSW code. 

\begin{figure}
    \centering
    \includegraphics[width=4in]{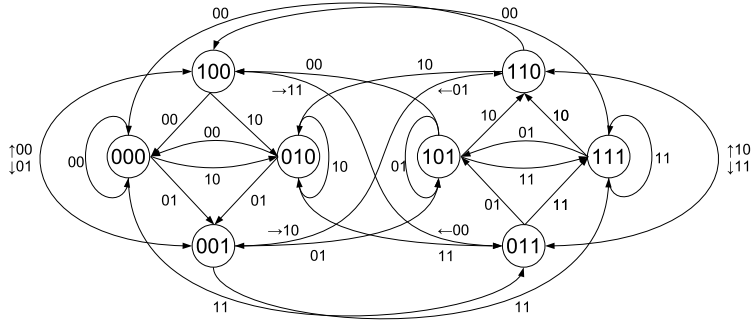}
    \caption{The FSM of the $(3, 2, 0)$-SSW code, in which each path denotes a valid walk of length $2$ in Fig.~\ref{fig:BruijnGraph}.}
    \label{fig:320SSWFSM}
\end{figure}

\begin{figure}
    \centering
    \includegraphics[width=3in]{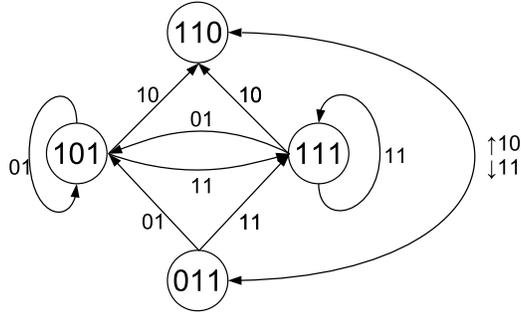}
    \caption{The FSM of the $(3, 2, 2)$-SSW code, which is a modified FSM of $(3, 2, 0)$-SSW code in Fig.~\ref{fig:320SSWFSM} by simply removing those vertices whose Hamming weight is less than $2$.}
    \label{fig:322SSWFSM}
\end{figure}

To obtain the FSM of the $(L, J, 0)$-SSW code with a skip length $J>1$, 
walks of length $J$ in the de Bruijn graph need to be extracted. 
An example of the FSM of the $(3, 2, 0)$-SSW code is depicted in Fig.~\ref{fig:320SSWFSM}.
Furthermore, letting $w([b_1\cdots b_L])=\sum_{i=1}^L b_i$ be the Hamming weight of the vertex $[b_1\cdots b_L]$ in $G(\setV , \setE )$,
the FSM of an $(L, J, 0)$-SSW code can be transformed into the FSM of an $(L, J, W)$-SSW code
by discarding vertices whose Hamming weights are less than $W$.
An example of the FSM of the $(3, 2, 2)$-SSW code is given in Fig.~\ref{fig:322SSWFSM}.

Based on the transformation of the FSM of the $(L, J, W)$-SSW code mentioned above, 
the adjacency matrix of the FSM corresponding to an $(L, J, W)$-SSW code can be derived as the following theorem. 
We use an operator that eliminates rows and columns of a matrix: $[\mB]_{\geq W}$ is defined as the submatrix 
of $\mB$ which deletes the rows and the columns of $\mB$ corresponding to those vertices whose Hamming weights are less than $W$.

\begin{lm}\label{lm:adjM}
Let $G(\setV , \setE )$ be the de Bruijn graph of order $L$ and the corresponding adjacency matrix be $\mA$. 
The FSM of the $(L, J, W)$-SSW code can be constructed as the adjacency matrix $[\mA^J]_{\geq W}$. 
\end{lm}
\begin{IEEEproof}
By \cite[Theorem 1.1]{Stanley2013}, the element at the $i$th row and $j$th column of the adjacency matrix $\mA^J$ 
is the number of valid walks from vertex $i$ to vertex $j$. 
Therefore, the adjacency matrix $\mA^J$ corresponds to the FSM of the $(L, J, 0)$-SSW code. 
As Fig.~\ref{fig:322SSWFSM} depicts, 
the FSM of the $(L, J, W)$-SSW code is the FSM of the $(L, J, 0)$-SSW code without those vertices whose Hamming weight is less than $W$.
Hence, the adjacency matrix of the $(L, J, W)$-SSW code can be obtained simply by removing the rows and columns of $\mA^J$
whose corresponding Hamming weight is less than $W$, i.e.\ $[\mA^J]_{\geq W}$.
\end{IEEEproof}

It should be noted that elements in matrix $[\mA^J]_{\geq W}$ are either $0$ or $1$ when $L\geq J$, 
and the size of the square matrix $[\mA^J]_{\geq W}$ is $b \times b$, where $b = \sum_{i=W}^{L}{L \choose i}$.
By Lemma \ref{lm:adjM}, the following theorem calculates the generating function \eqref{eqn:GF}.
\begin{thm}\label{thm:FSM}
The generating function of the $(L, J, W)$-SSW code is
\begin{equation*}
g(x)=1+\mOne^\mathrm{T} \bigg[ \mI-[\mA^J]_{\geq W}\cdot x^J\bigg]^{-1}\mOne x^L,
\end{equation*}
where $\mA$ is the adjacency matrix corresponding to the de Bruijn graph of order $L$.
\end{thm}
\begin{IEEEproof}
Letting $\mB=[\mA^J]_{\geq W}$ and $\mOne$ be the $b$-length column vector of ones,
the number of $(L, J, W)$-SSW sequences of length $L+kJ$ for some non-negative integer $k$ is
\begin{equation*}
M_\ssw^{(L, J, W)}(L+kJ)=\mOne^\mathrm{T}\mB^k\mOne.
\end{equation*}
Therefore, the generating function $g(x)$ can be derived as
\begin{IEEEeqnarray}{rCl}\label{eqn:FSM-gf}
g(x)&=&1+\sum_{k=0}^{\infty} \mOne^\mathrm{T}
\mB^k\mOne x^{L+kJ}\\\label{eqn:FSM-closed}
&=&1+\mOne^\mathrm{T} \left[ \mI-\mB x^J\right]^{-1}\mOne x^L,
\end{IEEEeqnarray}
where $\mI$ is the identity matrix of size $b\times b$.
\end{IEEEproof}

As per \cite[Lemma 3.5]{MarcusRS2001}, the logarithm of the largest absolute eigenvalue of $[\mA^J]_{\geq W}$ equals $J C_\ssw^{(L, J, W)}$:
\begin{IEEEeqnarray}{rCl}
C_\ssw^{(L, J,W)}&=& \lim_{k \to \infty} \frac{\log_2 \mOne^\mathrm{T}\big([\mA^J]_{\geq W}\big)^k \mOne}{L+kJ}\\
&=&\frac{\log_2 \lambda\big([\mA^J]_{\geq W}\big)}{J}, \label{eqn:capacity_compute}
\end{IEEEeqnarray}
where $\lambda\big([\mA^J]_{\geq W}\big)$ is the largest absolute value of all eigenvalues of $[\mA^J]_{\geq W}$.
Finding $C_\ssw^{(L, J, W)}$ is equivalent to finding the eigenvalue of the square matrix $[\mA^J]_{\geq W}$ 
of size $b\times b$.

\subsection{Goulden-Jackson Cluster Method with Bad Words}\label{sec:GJbad}
When the window length of the $(L, J, W)$-SSW sequences is a multiple of $J$, i.e., $L=\ell J$ for some positive integer $\ell$, 
we can apply Goulden-Jackson cluster method \cite{GouldenJ1979} to find the generating function $g(x)$.
The Goulden-Jackson cluster method is a technique to enumerate the valid sequences without any ``bad'' words within it. 
The Goulden-Jackson cluster method states that, 
given a set of letters $\setV$ and a set of bad words $\setB$ such that $\setB \in \setV^*$,
the generating function $f(x)$ for enumerating sequences containing no bad words within them can be expressed as
\begin{equation}\label{eqn:GJ-gf}
f(x)=\frac{1}{1-|\setV |x-\cluster_\setB (x)},
\end{equation}
where the $\cluster_\setB (x)$ is the generating function of the sequences of overlapped bad words. 
Since the sequences of overlapped bad words can be categorized by their last bad word,
$\cluster_\setB (x)$ can be computed by summing the generating function of all overlapped bad words ending with a different bad word, i.e.,
\begin{equation}\label{eqn:GJ-cluster}
\cluster_\setB (x)=\sum_{\bb\in \setB }\cluster_\setB (x|\bb) 
\end{equation} 
where $\cluster_\setB (x|\bb)$ is the generating function of the overlapped bad words ending with bad word $\bb$.
Based on Goulden-Jackson cluster method, 
the $\cluster_\setB (x|\bb) $ can be uniquely determined by solving the following $|\setB |$ linear equations:
\begin{equation}\label{eqn:GJ-ls}
\cluster_\setB (x|b) =
-x^{\Vert \bb\Vert}-\sum_{\bb' \in \setB }\sum_{\by\in\setO (\bb', \bb)} x^{\Vert\bb\Vert- \Vert\by\Vert}\cluster_\setB (x|\bb'), \mbox{ for all } \bb\in \setB ,
\end{equation}
where $\Vert\cdot\Vert$ denotes number of bits and 
\begin{equation}\label{eqn:GJ-setO}
\setO (\bb', \bb)\triangleq\left\{\by : \mbox{ there exist } \by, \bu, \bu'\in \{\setV^*\setminus \emptyset \} 
 \mbox{ such that } \bb' =\bu'\by \mbox{ and } \bb = \by\bu \right\}.
\end{equation}

Thus $(L, J, W)$-SSW sequences such that $L=\ell J$ for some positive integer $\ell$ 
can be enumerated by the Goulden-Jackson cluster method and its generating function can be calculated by the following theorem.
\begin{thm}\label{thm:GJ-SSW}
The generating function of the $(\ell J, J, W)$-SSW code for some positive integer $\ell$ is
\begin{equation}\label{eqn:GJ-SSW-gf}
g(x)=\frac{1}{1- |\setV| x^J-\cluster_\setB (x^J)},
\end{equation}
where $\setV=\{0,1\}^J$, $\setB=\{\bb: \bb\in\{0,1\}^{\ell J} \mbox{ and } w([\bb])<W\}$, $\cluster_\setB (x)=\sum_{\bb\in \setB }\cluster_\setB (x|\bb)$,
\begin{equation}\label{eqn:GJ-SSW-ls}
\cluster_\setB (x|b) =
-x^{\ell}-\sum_{\bb' \in \setB }\sum_{\by\in\setO (\bb', \bb)} x^{\ell- (\Vert\by\Vert/J)}\cluster_\setB (x|\bb'), \mbox{ for all } \bb\in \setB,
\end{equation}
and $\setO(\bb', \bb)$ is as in \eqref{eqn:GJ-setO}.
\end{thm}
\begin{IEEEproof}
Since the $(\ell J, J, W)$-SSW code is simply the language with alphabet $\setV$ such that no bad word in $\setB$ is included,
the Goulden-Jackson cluster method can be used directly.
Also, since $f(x)$ in \eqref{eqn:GJ-gf} enumerates the sequences from the alphabet $\setV$ instead of the binary alphabet, 
the generating function of $(\ell J, J, W)$-SSW sequences can be computed as 
$g(x)=f(x^J)$ which yields \eqref{eqn:GJ-SSW-gf}.
\end{IEEEproof}
Hence, finding $M_\ssw^{(\ell J, J, W)}(n)$ is equivalent to solving a linear system with $|\setB |=\sum_{i=0}^{W-1}{\ell J \choose i}$ unknowns in \eqref{eqn:GJ-SSW-ls}.

\subsection{Goulden-Jackson Cluster Method with Good Words}\label{sec:GJgood}
Conceptually, the FSM approach in Section \ref{sec:FSM} enumerates the $(L, J, W)$-SSW sequences by listing all legitimate sequences
whereas the Goulden-Jackson cluster method enumerates the $(\ell J, J, W)$-SSW sequences by excluding all invalid ones.  
The efficiency of the Goulden-Jackson cluster method for $(\ell J, J, W)$-SSW sequences depends on the number of bad words. 
The linear system \eqref{eqn:GJ-SSW-ls} is not easy to solve when $|\setB |=\sum_{i=0}^{W-1}{\ell J \choose i}$ is large. 
Borrowing from the FSM approach, the Goulden-Jackson cluster method with good words can be considered as an alternative, 
which computes the generating function by enumerating all valid sequences.
Ultimately, the Goulden-Jackson cluster method with good words 
converges to the FSM approach as the following theorem shows, 
providing a further interpretation of the Goulden-Jackson cluster method.

\begin{thm}\label{thm:FSMGJG}
To enumerate the $(\ell J, J, W)$-SSW sequences for some positive integer $\ell$, 
the FSM approach is equivalent to the Goulden-Jackson cluster method for enumerating overlapped good words.
\end{thm}
\begin{IEEEproof}
Let $\setV=\{0,1\}^J$ and $\setG$ be the set of good words, $\setG =\{0,1\}^{\ell J}\setminus \setB $.
Any $(\ell J, J, W)$-SSW sequence must be composed of consecutive good words and 
each good word must overlap with its neighbors with $(\ell-1)J$ bits. 
Hence, the generating function by Goulden-Jackson cluster method can be derived as
\begin{equation}\label{eqn:GJG-gf}
f(x)=1+\cluster_\setG (x),
\end{equation}
where $\cluster_\setG (x)=\sum_{\bb\in \setG } \cluster_\setG (x|\bb)$.
Similar to \eqref{eqn:GJ-SSW-ls}, $\cluster_\setG (x|\bb)$ for all $\bb\in \setG $ can be obtained by solving the following linear system
\begin{equation}\label{eqn:GJG-SSW-ls}
\cluster_\setG (x|\bb) =
x^\ell+\sum_{\bb' \in \setG }\sum_{\by\in\setO_\setG (\bb', \bb)} x\cdot \cluster_\setG (x|\bb'), \mbox{ for all } \bb\in \setG ,
\end{equation}
where
\begin{equation*}
\setO_\setG (\bb', \bb)\triangleq\left\{\by : \mbox{ there exist } \by\in \setV^{\ell -1}, \bu \mbox{ and } \bu'\in \setV 
\mbox{ such that } \bb' =\bu'\by \mbox{ and } \bb = \by\bu \right\}.
\end{equation*}
Since the linear system in \eqref{eqn:GJG-SSW-ls} can be rewritten in matrix form, 
\begin{equation*}
{\begin{bmatrix}
\cluster_\setG (x|\bb) 
\end{bmatrix}}_{\bb\in \setG }
=\mOne^{\mathrm{T}}x^\ell+\mB'x
{\begin{bmatrix}
\cluster_\setG (x|\bb) 
\end{bmatrix}}_{\bb\in \setG },
\end{equation*}
which can be solved as 
\begin{equation}\label{eqn:GJG-SSW-lssol}
{\begin{bmatrix}
\cluster_\setG (x|\bb) 
\end{bmatrix}}_{\bb\in \setG }
=\left[\mI-\mB' x\right]^{-1}\mOne x^\ell.
\end{equation}
Combining \eqref{eqn:GJG-gf} and \eqref{eqn:GJG-SSW-lssol}, 
\begin{equation*} 
g(x)=f(x^J)=1+\mOne^\mathrm{T} \left[ \mI-\mB' x^J\right]^{-1}\mOne x^{\ell J},
\end{equation*}
which coincides with \eqref{eqn:FSM-closed} since $\mB=\mB'$.
Therefore, FSM enumeration can be interpreted as the Goulden-Jackson cluster method for good words.
\end{IEEEproof}

\section{Refined Enumeration Methods}\label{sec:refined}
The methods proposed in the previous section find the generating function with computational complexity governed by the size of good word or bad word sets,
which can be exceedingly large in most practical cases. 
For example, to enumerate $(40, 20, 20)$-SSW sequences, 
the FSM approach must find eigenvalues of the square matrix of size $\sum_{i=20}^{40}{40 \choose i} \approx 6.2\times 10^{11}$ and
the Goulden-Jackson method needs to solve a linear system with $\sum_{i=0}^{19}{40 \choose i} \approx 4.8\times 10^{11}$ unknowns.
Here, we refine the methods of Section \ref{sec:enumeration} 
to reduce computational complexity.
Just a heads-up, in the case of enumerating $(40, 20, 20)$-SSW sequences, the refined FSM approach and the Goulden-Jackson method need to 
solve the linear systems with at most $21$ and $20$ unknowns, respectively.

\subsection{Refined Finite State Machine}
Considering the FSM $G(\setV , \setE )$ of the $(\ell J, J, W)$-SSW code for some positive integer $\ell$, 
each state $[\bb]=[b_1\cdots b_{\ell J}]\in \setV $ can be segmented into $\ell$ subblocks, i.e.,
$[\bb_k]=[b_{(k-1)J+1}b_{(k-1)J+2}\cdots b_{kJ}]$ for $1\leq k \leq \ell$.  
Let $w([\bb_k])$ be the Hamming weight of the subblock $[\bb_k]$, which is 
\begin{equation*}
w([\bb_k]) =w([b_{(k-1)J+1}\cdots b_{kJ}]) \mbox{, for all $1\leq k\leq \ell$}.
\end{equation*}
The following theorem shows that, 
for any two states $[\bb]$ and $[\bb']$ in $G(\setV , \setE )$ whose last $\ell-1$ subblocks have the same Hamming weight, respectively, 
i.e. $w([\bb_k])=w([\bb'_k])$ for all $2\leq k\leq \ell$, 
their outgoing edges are the same.
\begin{lm}\label{lm:RFSM}
Let $G(\setV , \setE )$ be the FSM of the $(\ell J, J, W)$-SSW code for some positive integer $\ell$ and let any two states $[\bb]$ and $[\bb']$ be in $\setV$.
If $w([\bb_k])=w([\bb'_k])$ for all $2\leq k\leq \ell$, then both edges
$\left( [\bb], [\bb_2\cdots \bb_\ell\, \by ] \right)$ and $\left( [\bb'], [\bb'_2 \cdots \bb'_\ell\, \by] \right)$ are in $\setE$
for all $\by\in \{0,1\}^J$ such that $w(\by) \geq W-w([\bb])+w([\bb_1])$. 
\end{lm}
\begin{IEEEproof}
Since both states $[\bb_2\cdots \bb_\ell\, \by ]$ and $[\bb'_2\cdots \bb'_\ell\, \by ]$ have the same Hamming weight: 
\begin{IEEEeqnarray}{rCl}
w([\bb_2\cdots \bb_\ell \by ])&=& w([\bb'_2\cdots \bb'_\ell \by ])\\
&=&\sum_{k=2}^{\ell}w([\bb_k])+w([\by])\\
&=&w([\bb])-w([\bb_1])+w([\by]) \geq W,
\end{IEEEeqnarray} 
and both edges $\left( [\bb], [\bb_2\cdots \bb_\ell \by ] \right)$ and $\left( [\bb'], [\bb'_2 \cdots \bb'_\ell \by] \right)$ are valid edges in $\setE$.
\end{IEEEproof}

By Lemma \ref{lm:RFSM}, the size of $G(\setV , \setE )$ 
can be reduced by grouping those states whose last $\ell-1$ subblocks have the same Hamming weights, respectively. 
Let $\bw=(w_1, w_2, \ldots, w_{\ell-1})$ be a vector of $\ell-1$ Hamming weights and define
$\setR (\bw)$ as
\begin{equation*}
\setR (\bw)\triangleq\Big\{ \bb=[b_1\cdots b_{\ell J}]: w([\bb])\geq W  \mbox{ and }
w([\bb_k])=w_{k-1} \mbox{ for all } 2\leq k\leq \ell \Big\},
\end{equation*}
which is the set of all valid states in $G(\setV , \setE )$ 
such that the Hamming weights of the last $\ell-1$ subblocks equal $\bw$.
The size of $\setR (\bw)$, denoted as $|\setR (\bw)|$, is
\begin{equation*}
|\setR (\bw)|=\left[\prod_{k=1}^{\ell -1} {J\choose w_k}\right]\times\left[\sum_{i=W-\sum_{j=1}^{\ell -1}w_j}^{J}{J \choose i}\right].
\end{equation*} 
We further define the set 
\begin{equation*}
\setW\triangleq\bigg\{ \bw: 0\leq w_k \leq \min\{J, W\} \mbox{ for all $1\leq k\leq \ell-1$, } 
\mbox{and }\sum_{i=1}^{\ell-1} w_i \geq W-J\bigg\},
\end{equation*} 
which is the set of all valid Hamming weight vectors for $(\ell J, J, W)$-SSW sequences.
The following theorem provides an efficient way to calculate the generating function \eqref{eqn:GF}.
\begin{thm}\label{thm:RFSM}
The generating function of the $(\ell J, J, W)$-SSW code for some positive integer $\ell$ is
\begin{equation}\label{eqn:RFSM-gf}
g(x)=1+\sum_{k=\ell}^{\infty} \mOne^\mathrm{T}\mA_\mathrm{R}^{k-\ell}\mv \times x^{kJ},
\end{equation}
where $\mA_\mathrm{R} =[m(\bw,\bw')]_{(\bw, \bw')\in \setW ^2}$ and $m(\bw, \bw')={J \choose w'_\ell}$.
\end{thm}
\begin{IEEEproof}
Any state in $\setR (\bw)$ can transit to another state in $\setR (\bw')$ 
if $w_k=w'_{k-1}$ for all $2\leq k\leq \ell-1$ and the edge can be represented as $J$ bits with $w'_\ell$ ones. 
Therefore, we can construct the reduced FSM $G_\mathrm{R} (\setV_\mathrm{R} , \setE_\mathrm{R} )$ as
\begin{equation*}
\setV_\mathrm{R} =\{\bw: \bw\in \setW \}
\end{equation*}
and 
\begin{equation*}
\setE_\mathrm{R} =\{(\bw , \bw' ): \bw, \bw' \in \setW  \mbox{ and }
w_k=w'_{k-1} \mbox{ for all } 2\leq k\leq \ell-1 \}.
\end{equation*}
The weight of edge $(\bw, \bw')\in \setE_\mathrm{R} $ is $m(\bw, \bw')={J \choose w'_\ell}$, 
which denotes the total number of possible transitions from vertex $\bw$ to vertex $\bw'$. 
\begin{figure}
    \centering
    \includegraphics[width=3in]{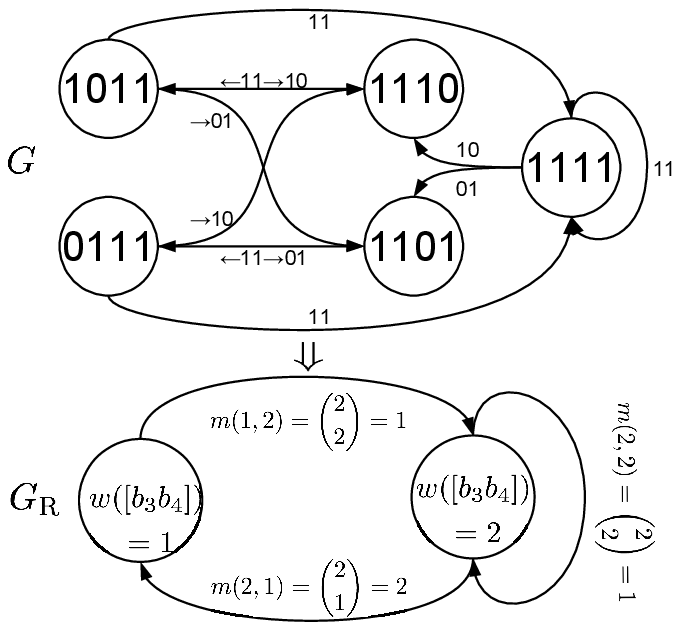}
    \caption{Converting the FSM $G$ of $(4, 2, 3)$-SSW code to its reduced FSM $G_{\mathrm{R}}$. 
    The weights of edges in $G_{\mathrm{R}}$ are given as $m([\bw, \bw'])$ for all $\bw, \bw'\in \setW$.}
    \label{fig:423SSWRFSM}
\end{figure}
Let $\mv$ be the column vector of $|\setR (\bw)|$ for all $\bw\in \setW $, which is 
\begin{equation*}
\mv=[|\setR (\bw)|]_{\bw\in \setW },
\end{equation*}
and let the adjacency matrix $\mA_\mathrm{R} $ of $G_\mathrm{R} (\setV_\mathrm{R} ,\setE_\mathrm{R} )$ be
\begin{equation*}
\mA_\mathrm{R} =[m(\bw,\bw')]_{(\bw, \bw')\in \setW ^2}.
\end{equation*}
Then
\begin{equation} \label{eqn:computeM}
M_\ssw^{(\ell J, J, W)}(kJ) = \mOne^{\mathrm{T}} \mA_\mathrm{R} ^{k-\ell} \mv ,  \\
\end{equation}
and the generating function can be derived as \eqref{eqn:RFSM-gf}.
\end{IEEEproof}

An example of converting the FSM of $(4, 2, 3)$-SSW code to its reduced FSM is given in Fig.~\ref{fig:423SSWRFSM},
where the reduced FSM is a weighted digraph. As shown, grouping all states in $G(\setV , \setE )$ 
yields a reduced FSM $G_\mathrm{R} (\setV_\mathrm{R} ,\setE_\mathrm{R} )$ with a size of $|\setW |$,
which is at most $\min\{J+1,W+1\}^{\ell-1}$. 
Hence, the $G_\mathrm{R} (\setV_\mathrm{R} ,\setE_\mathrm{R} )$ for the $(40, 20, 20)$-SSW code has at most $21$ states, 
a significant reduction compared to the $6.2\times 10^{11}$ states for $G(\setV , \setE )$. Further, using~\eqref{eqn:computeM}, the noiseless capacity for  skip-sliding window codes can efficiently be computed as
\begin{equation} \label{eqn:capacity_compute2}
C_\ssw^{(lJ, J,W)} = \frac{\log_2 \lambda\big( \mA_\mathrm{R} \big)}{J} , 
\end{equation}
where $\lambda\big( \mA_\mathrm{R} \big)$ is the largest positive eigenvalue of $\mA_\mathrm{R}$.

\subsection{Refined Goulden-Jackson Cluster Method with Bad Words}
Similar to the reduced FSM, we can apply the Goulden-Jackson cluster method to enumerate the $(\ell J, J, W)$-SSW sequences
by considering only the Hamming weight of the subblocks of each window.
The following theorem refines Theorem \ref{thm:GJ-SSW}.
\begin{thm}\label{thm:RGJ-SSW}
The generating function of the $(\ell J, J, W)$-SSW code for some positive integer $\ell$ is
\begin{equation}\label{eqn:RGJ-SSW-gf}
g(x)=\frac{1}{1-x^J \sum_{i=0}^{\min\{J,W-1\}}{J \choose i} -\cluster_{\setB} (x^J)},
\end{equation}
where $\setV =\big\{0, 1, \ldots , \min\{J,W-1\}\big\}$, 
$\setB=\big\{\bw:  \bw=w_1\cdots w_\ell\in \setV^{\ell-1} \mbox{ and } \sum_{i=1}^{\ell-1} w_i < W\big\}$,
$\cluster_\setB (x)=\sum_{\bw\in \setB }\cluster_\setB (x|\bw)$,
\begin{multline}\label{eqn:RGJ-SSW-ls}
\cluster_\setB (x|\bw) =
-x^{\ell}\left[\prod_{k=1}^{\ell-1} {J\choose w_k}\right]\times\left[\sum_{i=0}^{W-\sum_{j=1}^{\ell-1}w_j-1}{J \choose i}\right]\\
-\sum_{o=1}^{\ell-1} \Bigg\{ x^{\ell-o}\Bigg[\prod_{k=o}^{\ell-1}{J \choose w_k} \Bigg] \times 
\Bigg[ \sum_{i=0}^{W-\sum_{j=1}^{\ell-1}w_j-1}\sum_{\bw' \in \setO(i, o, \bw)}\cluster_\setB (x|\bw')\Bigg]\Bigg\},
\end{multline}
and 
\begin{equation}\label{eqn:RGJ-SSW-ol}
\setO(i, o, \bw)\triangleq\{\bw': \bw'\in \setB , w'_{\ell-o}=i \mbox{ and }\\
w'_{\ell-o+k}=w_{k} \mbox{ for all } 1 \leq k \leq o-1\}.
\end{equation}
\end{thm}
\begin{IEEEproof}
Similar to Theorem \ref{thm:RFSM}, a bit sequence of length $\ell J$ can be divided into $\ell$ subblocks 
and the Hamming weight of each subblock is at most $\min\{ J, W-1 \}$ to be a valid $(\ell J, J, W)$-SSW sequence. 
Hence, the $(\ell J, J, W)$-SSW code can be translated to be a language with alphabet $\setV$ such that
no sum of $\ell$ consecutive weights is less than $W$.
Since the number of binary sequences of length $J$ whose total Hamming weight is less than $W$ is $\sum_{i=0}^{\min\{J,W-1\}}{J \choose i}$, 
equation \eqref{eqn:GJ-gf} can be replaced by \eqref{eqn:RGJ-SSW-gf}.

As stated in the Goulden-Jackson cluster method, the $\cluster_\setB (x)$ is the generating function of the overlapped bad words
and $\cluster_\setB (x|\bw)$ denotes the generating function of those overlapped bad words that end with a bad word corresponding to $\bw\in\setB$. 
It should be noted that, for each $\ell$ consecutive weights, only the last $\ell -1$ weights $w_1\cdots w_{\ell-1}$ is used to denote the element in $\setB$
since the first weight can be any integer in $\Big\{0, 1, \ldots , W-\sum_{i=1}^{\ell-1}{w_i} -1\Big\}$. 
The number of binary sequences that are bad words of length $\ell J$ corresponding to $\bw=w_1\cdots w_{\ell-1}\in\setB$ 
can be computed as $\left[\prod_{k=1}^{\ell-1} {J\choose w_k}\right]\times\Big[\sum_{i=0}^{W-\sum_{j=1}^{\ell-1}w_j-1}{J \choose i}\Big]$, 
which is the coefficient of $-x^{\ell}$ in \eqref{eqn:RGJ-SSW-ls}. 
We can further enumerate those binary sequences which are overlapped by the bad words in the same way 
to derive the $\cluster_\setB (x|\bw)$ as in \eqref{eqn:RGJ-SSW-ls}.
\end{IEEEproof}

Hence, the refined Goulden-Jackson cluster method can enumerate the $(\ell J, J, W)$-SSW sequences 
by solving \eqref{eqn:RGJ-SSW-ls} with $|\setB|$ unknowns, where $|\setB |\leq \min\{J+1, W\}^{\ell-1}$.
For the example of $(40, 20, 20)$-SSW sequences, 
the refined Goulden-Jackson cluster method can find the generating function by solving the linear system with at most $20$ unknowns,
which is much smaller than the original $4.8\times 10^{11}$ unknowns.

Since Theorem \ref{thm:FSMGJG} shows the Goulden-Jackson cluster method for good words is an alternative interpretation of the FSM approach, 
and since the derivation of the refinement is similar to that of the bad words setting, it is omitted here.   

\section{Properties of $(L,J,W)$-SSW codes}\label{sec:properties}
In this section, we explore some properties of SSW codes. 
Let 
\begin{equation}\label{eqn:finiteC}
C_\ssw^{(L, J,W)}(L+kJ)\triangleq \frac{\log_2 M_\ssw^{(L,J, W)}(L+kJ)}{L+kJ},
\end{equation}
and let $S_\ssw^{(L, J,W)}(L+kJ)$ be the set of all $(L,J,W)$-SSW sequences of length $L+kJ$.
Two trivial inequalities of the noiseless capacity are given as the following lemmas, 
which show that the noiseless capacity increases as $W$ decreases and
the noiseless capacity decreases if the skip length $J$ is multiplied by a constant. 
\begin{lm}\label{lm-W}
Given positive integers $L$, $J$, $W$ and $W'$ such that $L>J$ and $L>W\geq W'>0$,
\begin{equation}
C_{\ssw}^{(L, J, W)}\leq C_{\ssw}^{(L,J, W')}.
\end{equation}
\end{lm}
\begin{IEEEproof}
Since any sequence in ($L, J, W$)-SSW is also in ($L, J, W'$)-SSW, $C_{\ssw}^{(L, J, W)}\leq C_{\ssw}^{(L,J, W')}$.
\end{IEEEproof}

\begin{lm}\label{lm-J}
Given positive integers $L$, $J$, $W$ and $k$ such that $L>kJ>0$,
\begin{equation}
C_{\ssw}^{(L, kJ, W)}\geq C_{\ssw}^{(L,J, W)}.
\end{equation}
\end{lm}
\begin{IEEEproof}
Since any sequence in ($L, J, W$)-SSW is also in ($L, kJ, W$)-SSW, $C_{\ssw}^{(L, kJ, W)}\geq C_{\ssw}^{(L,J, W)}$.
\end{IEEEproof}

We further examine finite blocklength properties in the noiseless case.
The following lemma shows the size of skip-sliding window codes can be upper-bounded by 
dividing into subblocks. 
\begin{lm}\label{lm-1}
Assume $L=\ell J$ for some integer $\ell>0$. Let $k, h \in \mathbb{Z}^{+}$ such that $h\geq \ell$ and $k\geq \ell $, then
\begin{equation}\label{eqn:lm-1-eq}
M_\ssw^{(L,J, W)}((h+k)J)\leq M_\ssw^{(L,J, W)}(hJ)\times M_\ssw^{(L,J, W)}(kJ),
\end{equation}
where equality holds if and only if $L=J$.
\end{lm}
\begin{IEEEproof}
Let 
\begin{equation*}
S'=\left\{\bb\bb': \bb\in S_\ssw^{(L, J,W)}(hJ), \bb'\in S_\ssw^{(L, J,W)}(kJ)\right\},
\end{equation*}
which is the set of all sequences which are the concatenations of any sequence in $S_\ssw^{(L, J,W)}(hJ)$ and
any sequence in $S_\ssw^{(L, J,W)}(kJ)$.
Since $S_\ssw^{(L, J,W)}((h+k)J)\subseteq S' $, 
\begin{IEEEeqnarray}{rCl}
M_\ssw^{(L, J,W)}((h+k)J)&\leq& |S'|\\
&=& M_\ssw^{(L,J, W)}(hJ)\times M_\ssw^{(L,J, W)}(kJ).
\end{IEEEeqnarray}
Since an $(L,J,W)$-SSW code reduces to an SEC code when $L=J$,
the equality of \eqref{eqn:lm-1-eq} always holds when $L=J$\cite{TandonMV2016, TandonKM2017}. 
Moreover, letting $\bb=1^{hJ-L+W}0^{L-W}$ and $\bb'=0^{L-W}1^{(kJ-L+W}$, 
it is clear that $\bb\in S_\ssw^{(L, J,W)}(hJ)$, $\bb' \in S_\ssw^{(L, J,W)}(kJ)$, but $\bb\bb' \notin S_\ssw^{(L, J,W)}((h+k)J)$ when $L > J$.
Hence the equality of \eqref{eqn:lm-1-eq} holds if and only if $L=J$.
\end{IEEEproof}

Lemma \ref{lm-1} provides a lower bound for $M_\ssw^{(L,J, W)}(hJ)\times M_\ssw^{(L,J, W)}(kJ)$. 
We further find an upper bound. 
\begin{lm}\label{lm-2}
Assume $L=\ell J$ for some integer $\ell>0$. Let $h,k \in \mathbb{Z}^{+}$ such that $h\geq \ell$ and $k\geq \ell$, then
\begin{equation}\label{eqn:lm-2-eq}
M_\ssw^{(L,J, W)}\left((h+k)J+(L-J)\right)\geq
M_\ssw^{(L,J, W)}(hJ)\times M_\ssw^{(L,J, W)}(kJ),
\end{equation}
where equality holds if and only if $L=W$.
\end{lm}
\begin{IEEEproof}
Let 
\begin{equation*}
S'=\left\{\bb1^{L-J}\bb': \bb \in S_\ssw^{(L, J,W)}(hJ)\mbox{ and }
\bb'\in S_\ssw^{(L, J,W)}(kJ)\right\},
\end{equation*}
Since $S'\subseteq S_\ssw^{(L, J,W)}\left((h+k)J+(L-J)\right)$, we have
\begin{IEEEeqnarray}{rCl}
M_\ssw^{(L,J, W)}\left((h+k)J+(L-J)\right)&\geq& |S'| \nonumber \\
&=&M_\ssw^{(L,J, W)}(hJ)\times M_\ssw^{(L,J, W)}(kJ).
\end{IEEEeqnarray}
When $L=W$, it is trivial that the equality of \eqref{eqn:lm-2-eq} holds since both sides equal $1$.
However, when $L>W$, a binary sequence $\bb=1^{hJ}01^{L-J-1}1^{kJ}$ is in $S_\ssw^{(L, J,W)}\left((h+k)J+(L-J)\right)$ but not in $S'$.
Hence the equality of \eqref{eqn:lm-2-eq} holds if and only if $L=W$.
\end{IEEEproof}

Let us consider the concatenation of subblocks with equal length. 
Lemmas~\ref{lm-1} and \ref{lm-2} can be extended to the following lemma. 

\begin{lm}\label{lm-3}
Assume $L=\ell J$ for some integer $\ell>0$. Let $h\in\mathbb{Z}^{+}$ such that $h\geq\ell$, then
\begin{equation}
M_\ssw^{(L, J,W)}(khJ)\leq \left[M_\ssw^{(L, J,W)}(hJ)\right]^k \leq M_\ssw^{(L, J,W)}\big(khJ+(k-1) (L-J)\big),
\end{equation}
for all integers $k>0$. The left equality holds if and only if $L=J$ and the right equality holds if and only if $L=W$.
\end{lm}
\begin{IEEEproof}
The proof is a direct extension of Lemmas \ref{lm-1} and \ref{lm-2}.
\end{IEEEproof}

To understand the properties of the noiseless capacity \eqref{eqn:capacity},
we further investigate properties of the rate as defined in \eqref{eqn:finiteC} when $L$ is a multiple of $J$.
Based on Lemma~\ref{lm-3},
the following lemma provides a lower bound and an upper bound on the rate.
\begin{lm}\label{lm-3-rate}
Assume $L=\ell J$ for some integer $\ell>0$.
Let $h \in \mathbb{Z}^{+}$ such that $h\geq \ell$, then
\begin{equation}\label{eqn:lm-3-eq}
C_\ssw^{(L,J, W)}(khJ)\leq C_\ssw^{(L,J, W)}(hJ)\leq \frac{kh+(k-1) (\ell-1)}{kh}C_\ssw^{(L, J,W)}\Big(\big[kh+(k-1) (\ell-1)\big]J\Big),
\end{equation}
for all integers $k>0$. The left equality holds if and only if $L=J$ and the right equality holds if and only if $L=W$.
\end{lm}
\begin{IEEEproof}
We first verify the left side of \eqref{eqn:lm-3-eq}. 
\begin{IEEEeqnarray}{rCl}
C_\ssw^{(L,J, W)}(khJ) &=& \frac{\log_2 M_\ssw^{(L, J,W)}(khJ)}{khJ}\\
&\leq& \frac{k \log_2 M_\ssw^{(L, J,W)}(hJ)}{khJ}\label{eqn:lm-3-leftineq}\\
&=&\frac{\log_2 M_\ssw^{(L, J,W)}(hJ)}{hJ}\\
&=&C_\ssw^{(L,J, W)}(hJ),
\end{IEEEeqnarray}
where equality of \eqref{eqn:lm-3-leftineq} holds if and only if $L=J$. Hence the left side of of \eqref{eqn:lm-3-eq} is proven. 
The right side of \eqref{eqn:lm-3-eq} is further verified as the following. 
\begin{IEEEeqnarray}{rCl}
C_\ssw^{(L,J, W)}(khJ) &=& \frac{1}{khJ}\log_2 M_\ssw^{(L, J,W)}(khJ)\\
&\leq&\frac{1}{khJ}\log_2 M_\ssw^{(L, J,W)}\big(khJ+(k-1) (L-J)\big) \label{eqn:lm-3-rightineq}\\
&=&\frac{1}{khJ}\log_2 M_\ssw^{(L, J,W)}\Big(\big[kh+(k-1) (\ell-1)\big]J\Big)\\
&=&\frac{kh+(k-1) (\ell-1)}{kh}C_\ssw^{(L, J,W)}\Big(\big[kh+(k-1) (\ell-1)\big]J\Big),
\end{IEEEeqnarray}
where equality of \eqref{eqn:lm-3-rightineq} holds if and only if $L=W$.
Hence the right side of \eqref{eqn:lm-3-eq} is proven.
\end{IEEEproof}

From Lemma \ref{lm-3-rate}, the left side of \eqref{eqn:lm-3-eq} indicates that 
the rate of SSW codes decreases when its length is multiplied by a constant.  
Moreover, the right side of \eqref{eqn:lm-3-eq} implies that, 
when $h\gg k$, the rate loss in multiplying the length by $k$ can be very small.
The following lemma further shows that the rate of the $(L,J,W)$-SSW code is lower-bounded by its capacity.
\begin{lm}\label{lm-4}
Assume $L=\ell J$ for some integer $\ell>0$, then
\begin{equation}\label{eqn:lm-4-eq}
C_\ssw^{(L,J, W)}(L+k J)\geq C_\ssw^{(L,J, W)},
\end{equation}
for all integers $k\geq 0$.
The equality of \eqref{eqn:lm-4-eq} holds if and only if $L=J$.
\end{lm}
\begin{IEEEproof}
By Lemma \ref{lm-3-rate}, 
\begin{IEEEeqnarray}{rCl}
C_\ssw^{(L,J, W)}(L+k J) &\geq & C_\ssw^{(L,J, W)}\left(2(L+k J)\right) \label{eqn:lm-4-proof1}\\
&\geq & \lim_{n\to\infty} C_\ssw^{(L,J, W)}\left(2^n(L+k J)\right)  \label{eqn:lm-4-proof2}\\
&=&C_\ssw^{(L,J, W)},
\end{IEEEeqnarray}
where the equalities of \eqref{eqn:lm-4-proof1} and \eqref{eqn:lm-4-proof2} hold if and only if $L=J$.
\end{IEEEproof}

The rate of convergence of $C_\ssw^{(L,J, W)}(L+k J)$ to $C_\ssw^{(L,J, W)}$ can be given as the following lemma. 
\begin{lm}\label{lm-rateconvergence}
Let $\mB$ be the matrix of size $b \times b$ associated to the $(L,J,W)$-SSW code.
The rate of the $(L,J,W)$-SSW code is upper-bounded as 
\begin{equation}\label{eqn:lm-rateconvergence}
C_\ssw^{(L,J, W)}(L+k J)< C_\ssw^{(L,J, W)}+\frac{\sigma_b(k)}{kJ}\log_2(2^b-1)
+\frac{\nu_b(k)}{kJ}\log_2 \bigg(\frac{(\mOne\mB\mOne)^b}{\mOne\mB^b\mOne}\bigg)
\end{equation}
for all $k>0$, where 
\begin{IEEEeqnarray}{rCl}
\sigma_b(k)&=&
\begin{cases}
\frac{1}{2}\left(\log_2 k+1\right)\left(\log_2 k+2\right) \hfill \mbox{ for $b=2$},\\
\frac{(b-1)^3}{(b-2)^2}k^{\frac{\log_2(b-1)}{\log_2 b}} \hfill \mbox{ for $b>2$},
\end{cases}\\
\nu_b(k)&=&
\begin{cases}
\log_2 k+1\hfill \mbox{ for $b=2$},\\
\frac{(b-1)^2}{(b-2)}k^{\frac{\log_2(b-1)}{\log_2 b}}\hfill \mbox{ for $b>2$}.
\end{cases}
\end{IEEEeqnarray}
\end{lm}
\begin{IEEEproof}
Let $\mB$ be the transition matrix corresponding to the $(L,J,W)$-SSW code.
\begin{IEEEeqnarray}{rCl}
C_\ssw^{(L,J, W)}&=&\lim_{k \to \infty} \frac{\log_2 \mOne^\mathrm{T}\big(\mB\big)^k \mOne}{L+kJ}\\
&\geq&\frac{\log_2 \mOne^\mathrm{T}\big(\mB\big)^k \mOne}{kJ}-\frac{\sigma_b(k)}{kJ}\log_2(2^b-1)
+\frac{\nu_b(k)}{kJ}\log_2 \bigg(\frac{(\mOne\mB\mOne)^b}{\mOne\mB^b\mOne}\bigg)\label{eqn:kozyakin}\\
&>&\frac{\log_2 \mOne^\mathrm{T}\big(\mB\big)^k \mOne}{L+kJ}-\frac{\sigma_b(k)}{kJ}\log_2(2^b-1)
+\frac{\nu_b(k)}{kJ}\log_2 \bigg(\frac{(\mOne\mB\mOne)^b}{\mOne\mB^b\mOne}\bigg)\\
&=&C_\ssw^{(L,J, W)}(L+k J)-\frac{\sigma_b(k)}{kJ}\log_2(2^b-1)
+\frac{\nu_b(k)}{kJ}\log_2 \bigg(\frac{(\mOne\mB\mOne)^b}{\mOne\mB^b\mOne}\bigg)
\end{IEEEeqnarray}
where inequality \eqref{eqn:kozyakin} is from \cite[Thm.~1]{Kozyakin2009}.
\end{IEEEproof}

Based on Lemmas \ref{lm-3-rate}, \ref{lm-4}, and \ref{lm-rateconvergence}, 
the rate of SSW codes seems to be non-increasing along the length when $L=\ell J$ for some $\ell\in\mathbb{Z}^{+}$
and it converges to its capacity asymptotically;
however, we have been unable to prove this. 
We leave the non-increasing properties as the following conjecture.
\begin{conj}\label{conj-1}
Assume $L=\ell J$ for some integer $\ell>0$. Let $h \in \mathbb{Z}^{+}$ such that $h\geq \ell$, 
then we conjecture that the rate of the SSW code decreases if the length of the code is multiplied by a positive integer as 
\begin{equation}\label{lm-5-inq}
C_\ssw^{(L,J, W)}((k+1)hL)\leq C_\ssw^{(L,J, W)}(khL),
\end{equation}
for all integers $k>0$.
\end{conj}
It should be noted that the conjecture is only stated for the case where $L$ is a multiple of $J$. In fact, the next section presents counterexamples where \eqref{lm-5-inq} does not hold when $L \neq \ell J$.

\emph{Remark:} The following stronger statement is not true, even when $L=\ell J$. For all integers $k\geq0$
\begin{equation} \label{lm-5-inq_v2}
C_\ssw^{(L,J, W)}(L+(k+1)J)\leq C_\ssw^{(L,J, W)}(L+kJ).
\end{equation}
This can be seen from a counterexample, i.e., 
for $k=2$, $C_\ssw^{(10,5, 9)}(25)=0.3293 > C_\ssw^{(10,5,9)}(20)=0.3292$.

\begin{figure}[t]
    \centering
    \includegraphics[width=5in]{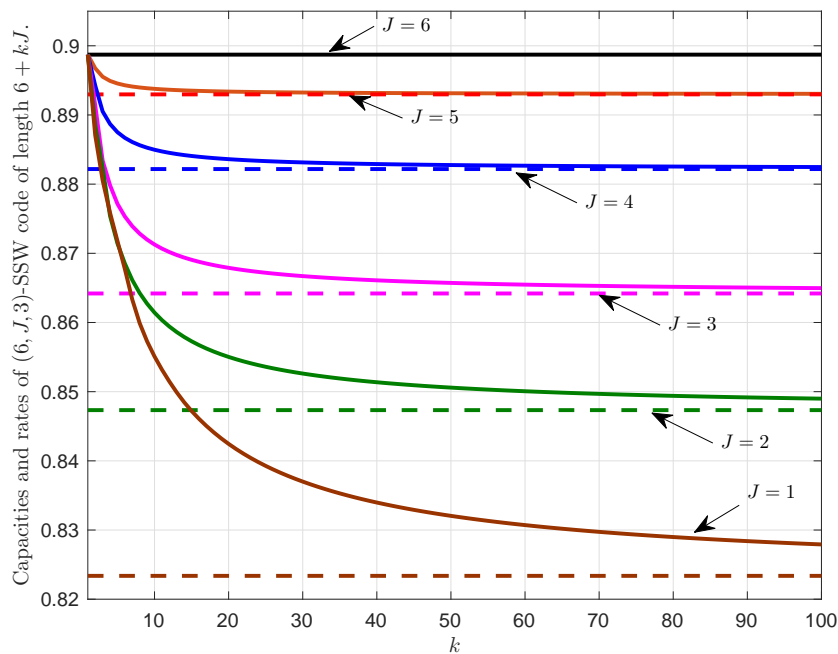}
    \caption{The capacities and rates of $(6,J,3)$-SSW codes for different $J$. 
    Capacities are drawn with dotted lines and rates with solid lines.}
    \label{fig:l6}
\end{figure}

\subsection{Numerical results}\label{sec:noieselessresult}
Here, some numerical computations are performed to give more insights into the performance of $(L, J, W)$-SSW codes in the noiseless case. 
Moreover, some counterintuitive observations are made.

Fig.~\ref{fig:l6} compares the capacities and the rates of $(6,J,3)$-SSW codes for different $J$, 
in which capacity is plotted as dotted lines and rate is plotted as solid lines. 
Since an $(L,J,W)$-SSW code reduces to an SEC code when $L=J$,
the curve of $J=6$ has a constant rate as a function of length 
because the rate of the SEC code does not depend on length \cite{TandonMV2016}. 
Some further remarks can be made from Fig.~\ref{fig:l6}. 
First, higher capacity can be achieved by lengthening the skip length $J$ for $(6,J,3)$-SSW codes, 
which coincides with the intuition that shorter $J$ will strengthen the sliding constraint
and the stronger constraint should lower the size of the code.  
Second, the rates of $(6,J,3)$-SSW codes seem to be non-increasing with length, 
which provide support for the conjecture in the previous section.

\begin{figure}
    \centering
    \includegraphics[width=5in]{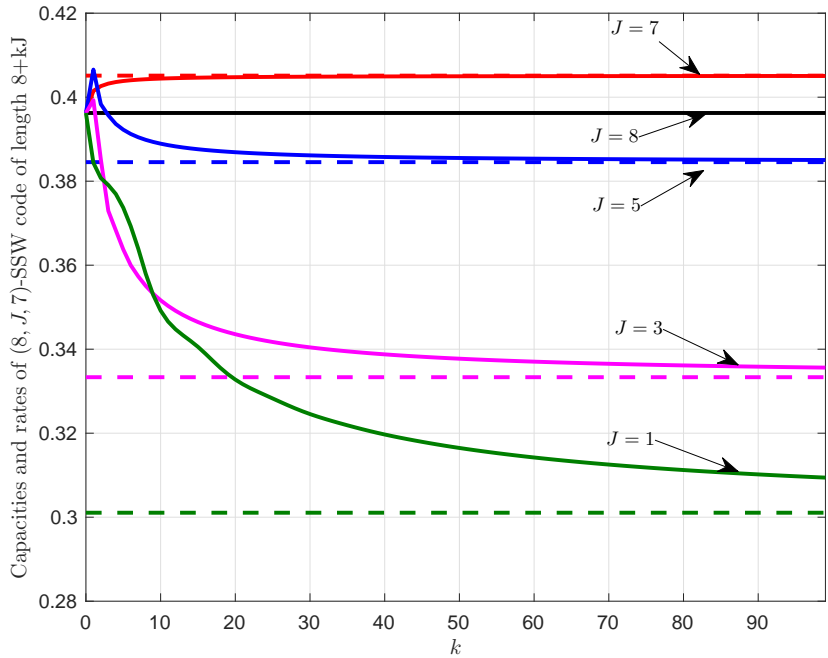}
    \caption{The capacities and rates of $(8,J,7)$-SSW codes for different $J$. 
    Capacities are drawn with dotted lines and rates with solid lines.}
    \label{fig:l8}
\end{figure}

The $(8,J,7)$-SSW codes, however, have completely different performance properties from $(6,J,3)$-SSW codes.
Similar to Fig.~\ref{fig:l6}, Fig.~\ref{fig:l8} plots capacities and rates for $(8,J,7)$-SSW codes.
We list some important points from Fig.~\ref{fig:l8} as follows.
First, the rates of $(8,J,7)$-SSW codes are no longer non-increasing as a function of the length. 
The $J=5$ and $J=3$ curves show that their rates are non-increasing except when $K$ is close to $1$.
In particular, the curve for $J=7$ even shows non-decreasing rate as a function of the length. 
Due to this observation, the case of $L\neq \ell J$ has been ruled out from Conjecture \ref{conj-1}.

Second, comparing the curves for $J=8$ and $J=7$, 
we surprisingly see that an $(L,J,W)$-SSW code with a longer $J$ does not guarantee a higher capacity, 
contrary to intuition. 
Fig.~\ref{fig:excc} had suggested that an $(L,J,W)$-SSW code with a shorter $J$ implies a stronger constraint is applied, 
which means the capacity of the $(L,J,W)$-SSW code should be higher than the $(L,J-1,W)$-SSW code. 
However, this numerical computation shows the contrary result and is useful in applications of SSW codes. 
For simultaneous information and energy transmission, this observation implies that 
some SSW codes can have higher capacity than SEC codes while also guaranteeing smoother energy transmission.
 
\section{Noisy Capacity bounds of Skip-Sliding Window Codes}\label{sec:noisy}
In this section, we present bounds on the noisy capacity of binary SSW codes.
In particular, we consider binary symmetric channels (BSCs) and binary erasure channels (BECs).

We first discuss lower bounds on the noisy capacity. Let $C_{\ssw,\,\BSC(p)}^{(L, J, W)}$ denote the capacity 
of SSW codes over a BSC with crossover probability $p$.
Then using Mrs. Gerber's Lemma (MGL)~\cite{WynerZ1973}, the noisy channel capacity $C_{\ssw,\,\BSC(p)}^{(L, J, W)}$ can be lower bounded as follows.
\begin{lm} \label{lem:BSC_MGL}
We have
\begin{equation} \label{eqn:capacity_BSC_MGL}
C_{\ssw,\,\BSC(p)}^{(L, J, W)} \ge h(\alpha \star p) - h(p) ,
\end{equation}	
where the binary operator $\star$ is defined as $a \star b \triangleq a(1-b) + (1-a)b$, ~$h(\cdot)$
is the binary entropy function, and $\alpha$ is chosen such that $h(\alpha) = C_{\ssw}^{(L, J, W)}$
with $0 \le \alpha \le 0.5$. The bound~\eqref{eqn:capacity_BSC_MGL} is tight for $p \to 0$.
\end{lm}
\begin{IEEEproof}
Let $X_1^n = (X_1, X_2, \ldots, X_n)$ denote a sequence satisfying the $(L,J,W)$-SSW constraint, and let
$Y_1^n = (Y_1, Y_2, \ldots, Y_n)$ be the corresponding output from a BSC with crossover probability $p$.
Then
\begin{equation*}
C_{\ssw,\,\BSC(p)}^{(L, J, W)} = \lim_{n \to \infty} \sup_{Q(X_1^n)} \frac{H(Y_1^n) - H(Y_1^n | X_1^n)}{n} ,
\end{equation*} 
where the supremum is taken over all input probability distributions $Q(X_1^n)$ of the sequences $X_1^n$ satisfying the
$(L,J,W)$-SSW constraint~\eqref{eqn:ssw}. A lower bound on $C_{\ssw,\,\BSC(p)}^{(L, J, W)}$ is obtained when $X_1^n$
is uniformly distributed over the set of $n$-length sequences satisfying the $(L,J,W)$-SSW constraint.
Thus, for uniformly distributed $X_1^n$, the constrained capacity over BSC can be lower bounded as follows.
\begin{align}
C_{\ssw,\,\BSC(p)}^{(L, J, W)} &\ge \lim_{n \to \infty} \frac{H(Y_1^n)}{n} - \frac{H(Y_1^n | X_1^n)}{n} \nonumber \\
&= \lim_{n \to \infty} \frac{H(Y_1^n)}{n} - H(Y_1 | X_1) \nonumber \\
&= \lim_{n \to \infty} \frac{H(Y_1^n)}{n} - h(p) . \label{eqn:BSC_cap1}
\end{align} 
Now, the noiseless capacity $C_\ssw^{(L, J, W)} = h(\alpha)$, denotes the entropy rate of a binary source which produces sequences
satisfying the $(L,J,W)$-SSW constraint, when the feasible input sequences are uniformly distributed.
When these constrained sequences are transmitted over a BSC with crossover probability $p$, then using Mrs. Gerber's Lemma (MGL),
the output entropy rate is lower bounded as~\cite{ZehaviW1988}
\begin{equation} \label{eqn:BSC_MGL}
\lim_{n \to \infty} \frac{H(Y_1^n)}{n} \ge h(\alpha \star p) , 
\end{equation}
and we obtain \eqref{eqn:capacity_BSC_MGL} by combining \eqref{eqn:BSC_cap1} and \eqref{eqn:BSC_MGL}.
The tightness of the lower bound~\eqref{eqn:capacity_BSC_MGL} follows as
\begin{equation*}
\lim_{p \to 0} C_{\ssw,\,\BSC(p)}^{(L, J, W)} = h(\alpha) = C_{\ssw}^{(L, J, W)} .
\end{equation*}  
\end{IEEEproof}

Let $C_{\ssw,\,\BEC(\epsilon)}^{(L, J, W)}$ denote the capacity of SSW codes
over a BEC with erasure probability~$\epsilon$.
Using an extension of MGL for binary input symmetric channels~\cite{ChayatS1989}, $C_{\ssw,\,\BEC(\epsilon)}^{(L, J, W)}$
can be lower bounded as follows.
\begin{lm} \label{lem:BEC_MGL}
We have
\begin{equation} \label{eqn:capacity_BEC_MGL}
C_{\ssw,\,\BEC(\epsilon)}^{(L, J, W)} \ge (1-\epsilon) C_\ssw^{(L, J, W)} ,
\end{equation}	
and this bound is tight for $\epsilon \to 0$.
\end{lm}
\begin{IEEEproof}
Let $X_1^n = (X_1, X_2, \ldots, X_n)$ denote a sequence satisfying the $(L,J,W)$-SSW constraint, and let
$Y_1^n = (Y_1, Y_2, \ldots, Y_n)$ be the corresponding output from a BEC with erasure probability $\epsilon$.
Then,
\begin{align*}
C_{\ssw,\,\BEC(\epsilon)}^{(L, J, W)} &= \lim_{n \to \infty} \sup_{Q(X_1^n)} \frac{H(Y_1^n)}{n} - \frac{H(Y_1^n | X_1^n)}{n} \\
&=  \lim_{n \to \infty} \sup_{Q(X_1^n)} \frac{H(Y_1^n)}{n} - h(\epsilon) , 
\end{align*} 
where the supremum is taken over all input probability distributions $Q(X_1^n)$ of the sequences $X_1^n$ satisfying the
$(L,J,W)$-SSW constraint. When $X_1^n$ is uniformly distributed over the set of feasible input sequences, we get
\begin{equation} \label{eqn:BEC_cap1}
C_{\ssw,\,\BEC(\epsilon)}^{(L, J, W)} \ge \lim_{n \to \infty} \frac{H(Y_1^n)}{n} - h(\epsilon) , 
\end{equation} 

Now, the noiseless capacity $C_\ssw^{(L, J, W)}$ denotes the entropy rate of a binary source which produces sequences
satisfying the $(L,J,W)$-SSW constraint, when feasible input sequences are uniformly distributed. When these constrained sequences are 
transmitted over a BEC with erasure probability $\epsilon$, then using an extension of MGL for binary-input symmetric channels,
the output entropy rate is lower bounded as~\cite{ChayatS1989}
\begin{equation} \label{eqn:BEC_MGL}
\lim_{n \to \infty} \frac{H(Y_1^n)}{n} \ge (1-\epsilon) C_\ssw^{(L, J, W)} + h(\epsilon) , 
\end{equation}
and we obtain \eqref{eqn:capacity_BEC_MGL} by combining \eqref{eqn:BEC_cap1} and \eqref{eqn:BEC_MGL}.
The tightness of the bound follows as
\begin{equation*}
\lim_{\epsilon \to 0} C_{\ssw,\,\BEC(\epsilon)}^{(L, J, W)} = C_{\ssw}^{(L, J, W)} .
\end{equation*}  
\end{IEEEproof}

\emph{Remark:}~The noiseless capacity provides a trivial upper bound on the capacity of noisy channels. Therefore, Lemmma~\ref{lem:BSC_MGL} (resp. Lemma~\ref{lem:BEC_MGL}) implies that if the noiseless capacity of two different SSW codes satisfies $C_\ssw^{(L_1, J_1, W_1)} > C_\ssw^{(L_2, J_2, W_2)}$, then for sufficiently small crossover probability $p$ (resp. erasure probability $\epsilon$), we have the inequality $C_{\ssw,\,\BSC(p)}^{(L_1, J_1, W_1)} > C_{\ssw,\,\BSC(p)}^{(L_2, J_2, W_2)}$ \Big(resp. $C_{\ssw,\,\BEC(\epsilon)}^{(L_1, J_1, W_1)} > C_{\ssw,\,\BEC(\epsilon)}^{(L_2, J_2, W_2)}$\Big).

Alternate lower bounds on the noisy capacity of skip-sliding window codes can be obtained using a generic
bound by Zehavi and Wolf~\cite[Lemma 4]{ZehaviW1988} on the noisy capacity of constrained sequences generated by a
Markov source. As shown in Sec.~\ref{sec:FSM}, an $(L,J,W)$-SSW code can be represented via a finite state
machine (FSM).  Thus, a source producing SSW constrained sequences can be modeled as a stationary
Markov source with non-zero probabilities associated with feasible state transitions in the corresponding
FSM. 

As shown in Sec.~\ref{sec:FSM}, a binary $(L,J,W)$-SSW code forms a FSM, in which 
$J$ consecutive uses of the channel can be viewed as a single use of a \emph{vector
channel} with \emph{super-letter input alphabet} $\setX=\{0,1\}^J$ and \emph{super-letter output alphabet} $\setY=\{0,1\}^J$. Let $\setS = \{\bs_1, \bs_2, \ldots, \bs_k\}$ be the set of $k$ distinct states in a FSM associated with the corresponding $(L ,J,W)$-SSW code. For $1 \le i, j \le k$, let $q_{i,j}$ be the probability that FSM transitions to state
$\bs_j$, given that the current state is $\bs_i$. Let $\bQ = [q_{i,j}]$ denote the
state-transition probability matrix, and let $\bx_{ij}$ be the super-letter symbol produced when FSM transitions from state $\bs_i$ to $\bs_j$. Further, let $Pr(\bS = \bs_i)$ denotes the steady-state probability that FSM will be in state $\bs_i$. Then the capacity $C_{\ssw,\,\BSC(p)}^{(L, J, W)}$ over BSC with crossover probability $p$ is lower bounded as follows.
\begin{lm} \label{lem:BSC_ZW}
	We have
	\begin{equation} \label{eqn:capacity_BSC_ZW}
	C_{\ssw,\,\BSC(p)}^{(L, J, W)} \ge \sup_{\bQ} ~\sum_{i=1}^k Pr(\bS = \bs_i) \frac{H(\bY | \bS = \bs_i)}{J} \, - \, h(p) , 
	\end{equation}
	where the conditional distribution for output super-letter $\bY$, is given by 
	\begin{align} 
	Pr(\by | \bS = \bs_i) &= \sum_{j=1}^{k} q_{i,j} Pr(\by | \bx_{ij}) , \ \ \ \ \ \ \ \ \ \ \ \  1 \le i \le k, \ \bx_{ij} \in \setX^J, \by \in \setY^J  \nonumber \\
	Pr(\by | \bx_{ij}) &= p^{d(\by,\bx_{ij})} (1-p)^{J - d(\by,\bx_{ij})} , 
	\end{align}
	with $\setX = \setY = \{0,1\}$ and $d(\by,\bx_{ij})$ denotes the Hamming distance between super-letters $\by$ and $\bx_{ij}$.
\end{lm} 
\begin{IEEEproof}
For a given $(L, J, W)$-SSW constraint, consider the corresponding FSM with state space $\setS = \{\bs_1, \bs_2, \ldots, \bs_k\}$. The transition probability from state $\bs_i$ to state $\bs_j$ is $q_{i,j}$. Let $\bS$ denote the previous state, and let $\tilde{\bS}$ denote the current state. When $2^J$-ary super-letters produced from this Markov source are transmitted over a memoryless channel, then the super-letter capacity is lower bounded by the conditional mutual information term~\cite[Lemma 4]{ZehaviW1988} $\sup_{Pr(\bS,\tilde{\bS})} I(\bY,\tilde{\bS} | \bS)$. Because the super-letter capacity corresponds to $J$ uses of the channel, the scalar capacity per channel use can therefore be lower bounded as follows.
\begin{align*}
	C_{\ssw,\,\BSC(p)}^{(L, J, W)} &\ge \sup_{Pr(\bS,\tilde{\bS})} \frac{I(\bY,\tilde{\bS} | \bS)}{J} \\
	&= \sup_{Pr(\bS,\tilde{\bS})} \left( \frac{H(\bY | \bS)}{J} - \frac{H(\bY | \bS , \tilde{\bS})}{J} \right) \\
	&= \sup_{\bQ} ~\sum_{i=1}^k Pr(\bS = \bs_i) \left(\frac{H(\bY | \bS = \bs_i)}{J} - \sum_{j=1}^k q_{i,j}\,\frac{H(\bY | \bX = \bx_{ij})}{J} \right) \\
	&\overset{(\mathrm{i})}{=} \sup_{\bQ} ~\sum_{i=1}^k Pr(\bS = \bs_i) \left(\frac{H(\bY | \bS = \bs_i)}{J} - \sum_{j=1}^k q_{i,j} \, h(p) \right) \\
	&= \sup_{\bQ} ~\sum_{i=1}^k Pr(\bS = \bs_i) \frac{H(\bY | \bS = \bs_i)}{J} \, - \, h(p) ,
\end{align*}
where $(\mathrm{i})$ follows from the memoryless property of BSC.
\end{IEEEproof}

Now, consider a BEC with input alphabet $\setX = \{0,1\}$, output alphabet $\setY = \{0,1,e\}$, where $e$ denotes the erasure symbol and let the erasure probability be denoted~$\epsilon$. Then $J$ consecutive uses of this BEC will induce a vector-channel with input super-letter alphabet $\setX^J$ and output super-letter alphabet $\setY^J$. Further, for $1 \le m \le J$, let $x_{ij}^{(m)}$ denote the $m$-th letter of super-letter $\bx_{ij} \in \setX^J$ with $\bx_{ij} = (x_{ij}^{(1)} x_{ij}^{(2)} \ldots x_{ij}^{(J)})$. Similarly, for $\by \in \setY^J$, let $\by = (y^{(1)} y^{(2)} \ldots y^{(J)})$. Then, for this induced vector channel, the probability of receiving super-letter $\by \in \setY^J$, given that super-letter $\bx_{ij} \in \setX^J$ is transmitted, is given by
\begin{equation} \label{eqn:BEC_superletter_transition}
Pr(\by | \bx_{ij}) = 
\begin{cases}
0, ~~&\mathrm{if~} y^{(m)} \neq x_{ij}^{(m)} \mathrm{~and~} y^{(m)} \neq e, \mathrm{~~~for~} 1 \le m \le J, \\
\epsilon^{t_{\by}}\, (1-\epsilon)^{J-t_{\by}}, &\mathrm{otherwise} ,
\end{cases}
\end{equation}  
where $t_{\by}$ denotes the number of erasure symbols in output super-letter $\by$. The following lemma provides a lower bound to the capacity of $(L,J,W)$-SSW codes over BEC with erasure probability $\epsilon$.
\begin{lm} \label{lem:BEC_ZW}
	We have
	\begin{equation} \label{eqn:capacity_BEC_ZW}
	C_{\ssw,\,\BEC(\epsilon)}^{(L, J, W)} \ge \sup_{\bQ} ~\sum_{i=1}^k Pr(\bS = \bs_i) \frac{H(\bY | \bS = \bs_i)}{J} \, - \, h(\epsilon) , 
	\end{equation}
	where the distribution for output super-letter $\bY$, given that the transmitted input super-letter is $\bX = \bx_{ij}$, is given by~\eqref{eqn:BEC_superletter_transition}, and  $Pr(\by | \bS = \bs_i) = \sum_{j=1}^k q_{i,j} \, Pr(\by | \bx_{ij})$.	
\end{lm}
The proof of the above lemma follows using steps similar to those used in the proof of Lemma~\ref{lem:BSC_ZW}, and is hence omitted.

We now provide upper bounds to the noisy capacity of $(\ell J,J,W)$-SSW codes.
\begin{lm} \label{lem:BSC_UB}
	We have
	\begin{equation} \label{eqn:capacity_BSC_UB}
	C_{\ssw,\,\BSC(p)}^{(\ell J, J, W)} \le \min\left\{C_{\ssw}^{(\ell J, J, W)}, C_{\ssw,\,\BSC(p)}^{(\ell J, \ell J, W)}\right\} . 
	\end{equation}
\end{lm} 
\begin{IEEEproof}
The noisy capacity $C_{\ssw,\,\BSC(p)}^{(\ell J, J, W)}$ is obviously upper bounded by the noiseless capacity $C_{\ssw}^{(\ell J, J, W)}$. Further, as every $(\ell J,J,W)$-SSW sequence is also an $(\ell J,\ell J,W)$-SSW sequence, we have the inequality $C_{\ssw,\,\BSC(p)}^{(\ell J, J, W)} \le C_{\ssw,\,\BSC(p)}^{(\ell J, \ell J, W)}$.	
\end{IEEEproof}
Note that $C_{\ssw,\,\BSC(p)}^{(\ell J, \ell J, W)}$ corresponds to the capacity of subblock energy-constrained (SEC) codes with subblock length $\ell J$ and subblock weight at least $W$, over a BSC with crossover probability $p$. This capacity term can numerically be computed for reasonable subblock lengths using the Blahut-Arimoto algorithm~\cite{Blahut1972,Arimoto1972}, by applying the super-letter approach for characterizing the capacity of SEC codes over arbitrary discrete memoryless channels in~\cite{TandonMV2016}. 

Similar to Lemma~\ref{lem:BSC_UB}, the following lemma provides an upper bound to the capacity of $(\ell J,J,W)$-SSW codes over BEC.
\begin{lm} \label{lem:BEC_UB}
	We have
	\begin{equation} \label{eqn:capacity_BEC_UB}
	C_{\ssw,\,\BEC(\epsilon)}^{(\ell J, J, W)} \le \min\left\{C_{\ssw}^{(\ell J, J, W)}, C_{\ssw,\,\BEC(\epsilon)}^{(\ell J, \ell J, W)}\right\} . 
	\end{equation}
\end{lm} 

\subsection{Numerical Results}

\begin{figure}
	\centering
	\includegraphics[width=5in]{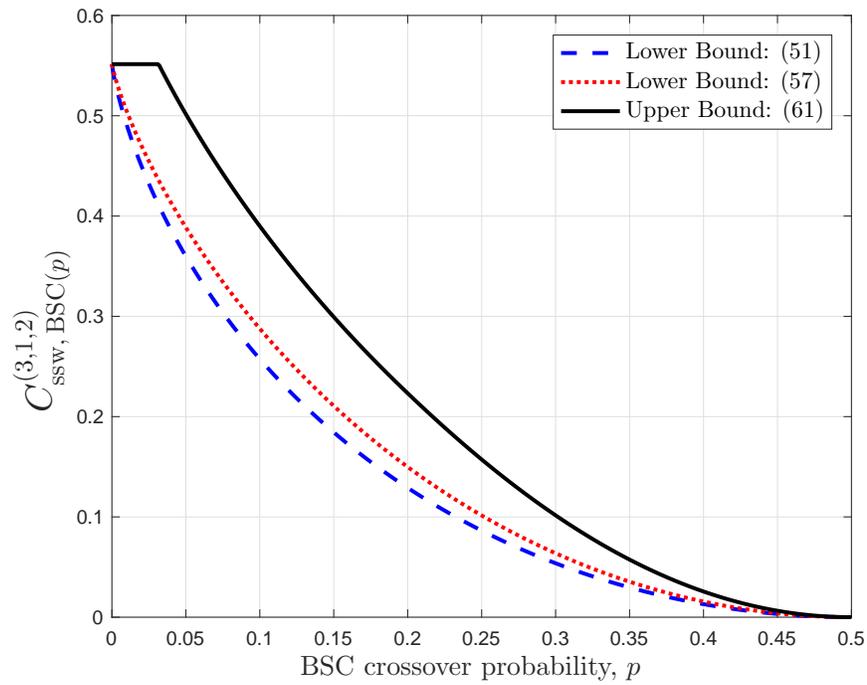}
	\caption{Bounds on the noisy channel capacity of $(3,1,2)$-SSW codes over BSC.}
	\label{fig:SSW312_BSC}
\end{figure}

\begin{figure}
	\centering
	\includegraphics[width=5in]{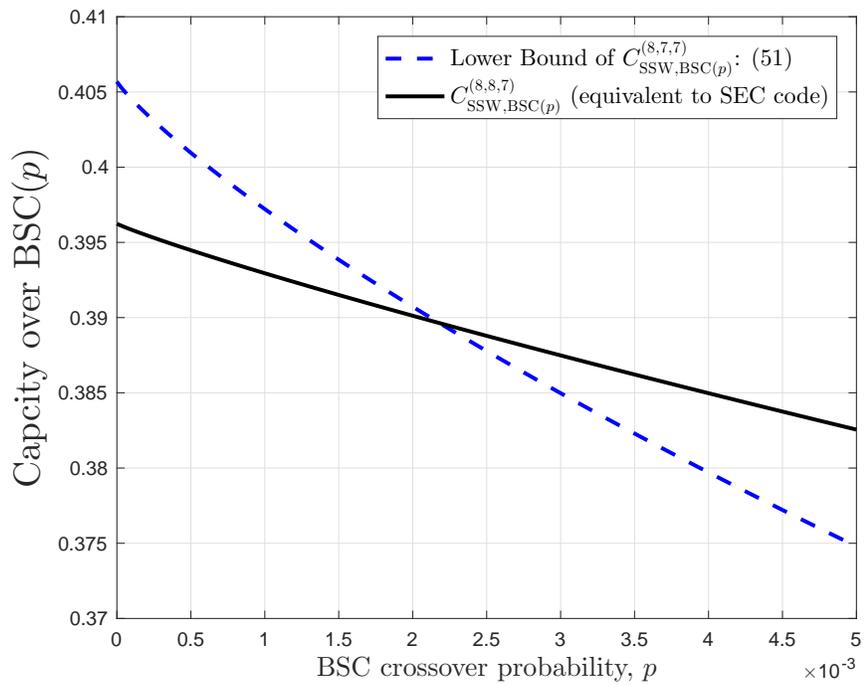}
	\caption{Comparison between the lower bound of $C_{\ssw,\,\BSC(p)}^{(8, 7, 7)}$ and the exact $C_{\ssw,\,\BSC(p)}^{(8, 8, 7)}$. 
	The latter one is also an SEC code.}
	\label{fig:SSW877_BSC}
\end{figure}

\begin{figure}
	\centering
	\includegraphics[width=5in]{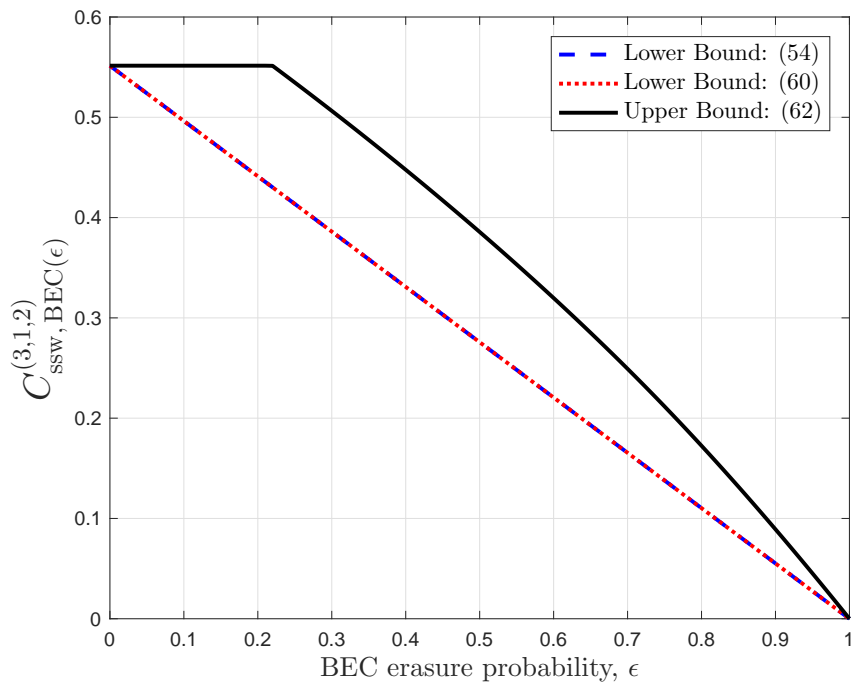}
	\caption{Bounds on the noisy channel capacity of $(3,1,2)$-SSW codes over BEC.}
	\label{fig:SSW312_BEC}
\end{figure}

In this subsection, we plot bounds on the noisy channel capacity of SSW codes. 
Fig.~\ref{fig:SSW312_BSC} plots bounds on capacity of $(L,J,W)$-SSW codes over a BSC with crossover probability $p$, for $L=3$, $J=1$, and $W=2$. The blue dotted line plots the lower bound on the noisy capacity given by~\eqref{eqn:capacity_BSC_MGL}, while the red line depicts the lower bound given by~\eqref{eqn:capacity_BSC_ZW}. In this case, it is seen that the lower bound~\eqref{eqn:capacity_BSC_ZW} is uniformly better than the lower bound obtained using Mrs. Gerber's Lemma~(MGL) in~\eqref{eqn:capacity_BSC_MGL}. The black curve, providing an upper bound on the capacity, is obtained using~\eqref{eqn:capacity_BSC_UB}. Further, it is seen that the bounds are tight for cases when $p \to 0$ and $p \to 0.5$.

We further examine the surprising observation from Fig.~\ref{fig:l8} that SSW codes may outperform SEC codes.
Fig.~\ref{fig:SSW877_BSC} compares the $(8,7,7)$-SSW code and $(8,8,7)$-SSW code over a BSC with a small crossover probability ($0\leq p \leq 0.005$). Here, the SEC capacity $C_{\ssw,\,\BSC(p)}^{(8, 8, 7)}$ is numerically computed~\cite{TandonMV2016} using the Blahut-Arimoto algorithm. For the $(8,7,7)$-SSW code, the capacity lower bound is plotted using~\eqref{eqn:capacity_BSC_MGL}.
It can be seen that the noisy capacity of SSW codes may be larger than that of SEC codes when $p$ is small, just like the noiseless case.

Fig.~\ref{fig:SSW312_BEC} plots bounds on capacity of $(3,1,2)$-SSW codes over a BEC with erasure probability $\epsilon$. The lower bounds on the capacity are given by~\eqref{eqn:capacity_BEC_MGL} and \eqref{eqn:capacity_BEC_ZW}, while the upper bound is obtained using~\eqref{eqn:capacity_BEC_UB}. In this case, it is observed that the two lower bounds coincide, and the bounds become tight for $\epsilon \to 0$ and $\epsilon \to 1$.

\section{Conclusion}\label{sec:con}
This paper proposed a new kind of constrained code, the skip-sliding window code,
which is potentially useful in diverse applications.
Efficient enumeration methods were proposed to calculate its noiseless capacity
and properties were discussed. 
Numerical results showed counterintuitive performance characterizations, 
such as the fact there can be skip-sliding window codes that outperform subblock-energy constraint codes \cite{TandonMV2016, TandonKM2017}
in both the capacity and the smoothness of energy transmission in simultaneous information and energy transmission.
Based on the numerical results, a conjecture on the noiseless capacity was also raised. 
With the help of noiseless capacity results, some noisy capacity bounds were further investigated; counterintuitive ordering of performance still holds.
\bibliographystyle{IEEEtran} 
\bibliography{full,conf_full,lrv_lib}

\newcommand{\SortNoop}[1]{}
\begin{thebibliography}{10}
\providecommand{\url}[1]{#1}
\csname url@samestyle\endcsname
\providecommand{\newblock}{\relax}
\providecommand{\bibinfo}[2]{#2}
\providecommand{\BIBentrySTDinterwordspacing}{\spaceskip=0pt\relax}
\providecommand{\BIBentryALTinterwordstretchfactor}{4}
\providecommand{\BIBentryALTinterwordspacing}{\spaceskip=\fontdimen2\font plus
\BIBentryALTinterwordstretchfactor\fontdimen3\font minus
  \fontdimen4\font\relax}
\providecommand{\BIBforeignlanguage}[2]{{%
\expandafter\ifx\csname l@#1\endcsname\relax
\typeout{** WARNING: IEEEtran.bst: No hyphenation pattern has been}%
\typeout{** loaded for the language `#1'. Using the pattern for}%
\typeout{** the default language instead.}%
\else
\language=\csname l@#1\endcsname
\fi
#2}}
\providecommand{\BIBdecl}{\relax}
\BIBdecl

\bibitem{Immink1990}
K.~A.~S. Immink, ``Runlength-limited sequences,'' \emph{Proceedings of the
  {IEEE}}, vol.~78, no.~11, pp. 1745--1759, Nov. 1990.

\bibitem{Immink2004}
------, \emph{Codes for Mass Data Storage Systems}.\hskip 1em plus 0.5em minus
  0.4em\relax Shannon Foundation Publisher, 2004.

\bibitem{UlukusYESZGH2015}
S.~Ulukus, A.~Yener, E.~Erkip, O.~Simeone, M.~Zorzi, P.~Grover, and K.~Huang,
  ``Energy harvesting wireless communications: a review of recent advances,''
  \emph{{IEEE} Journal on Selected Areas in Communications}, vol.~33, no.~3,
  pp. 360--381, Mar. 2015.

\bibitem{RosnesBY2012}
E.~Rosnes, {\'{A}}.~I. Barbero, and {\O}.~Ytrehus, ``Coding for inductively
  coupled channels,'' \emph{{IEEE} Transactions on Information Theory},
  vol.~58, no.~8, pp. 5418--5436, Aug. 2012.

\bibitem{FouladgarSE2014}
A.~M. Fouladgar, O.~Simeone, and E.~Erkip, ``Constrained codes for joint energy
  and information transfer,'' \emph{{IEEE} Transactions on Communications},
  vol.~62, no.~6, pp. 2121--2131, Jun. 2014.

\bibitem{TandonMV2014a}
A.~Tandon, M.~Motani, and L.~R. Varshney, ``On code design for simultaneous
  energy and information transfer,'' in \emph{Proceedings of the 2014
  Information Theory and Applications Workshop}, Feb. 2014.

\bibitem{TandonMV2016}
------, ``Subblock-constrained codes for real-time simultaneous energy and
  information transfer,'' \emph{{IEEE} Transactions on Information Theory},
  vol.~62, no.~7, pp. 4212--4227, Jul. 2016.

\bibitem{TandonKM2017}
A.~Tandon, H.~M. Kiah, and M.~Motani, ``Binary subblock energy-constrained
  codes: Bounds on code size and asymptotic rate,'' in \emph{Proceedings of the
  2017 IEEE International Symposium on Information Theory}, Jun. 2017, pp.
  1480--1484.

\bibitem{CohenST2010}
G.~Cohen, P.~Sol{\'{e}}, and A.~Tchamkerten, ``Heavy weight codes,'' in
  \emph{Proceedings of the 2010 IEEE International Symposium on Information
  Theory}, Jun. 2010, pp. 1120--1124.

\bibitem{BachocCCST2011}
C.~Bachoc, V.~Chandar, G.~Cohen, P.~Sol{\'{e}}, and A.~Tchamkerten, ``On
  bounded weight codes,'' \emph{{IEEE} Transactions on Information Theory},
  vol.~57, no.~10, pp. 6780--6787, Jul. 2011.

\bibitem{CompeauPT2011}
P.~E.~C. Compeau, P.~A. Pevzner, and G.~Tesler, ``How to apply de {B}ruijn
  graphs to genome assembly,'' \emph{Nature Biotechnology}, vol.~29, no.~11,
  pp. 987--991, Nov. 2011.

\bibitem{Bruijn1946}
N.~G. De~Bruijn, ``A combinatorial problem,'' \emph{Proceedings of the Section
  of Sciences}, vol.~49, no.~7, pp. 758--764, Jun. 1946.

\bibitem{BeldiceanuD2010}
N.~Beldiceanu and S.~Demassey, ``Sliding time window sum,'' Jan. 2010.

\bibitem{Salimov2010}
P.~V. Salimov, ``On {R}auzy graph sequences of infinite words,'' \emph{Journal
  of Applied and Industrial Mathematics}, vol.~4, no.~1, pp. 127--135, Jan.
  2010.

\bibitem{Rigo2014}
M.~Rigo, \emph{Formal Languages, Automata and Numeration Systems}.\hskip 1em
  plus 0.5em minus 0.4em\relax New York: John Wiley \& Sons, 2014.

\bibitem{HoevePRS2006}
W.-J. van Hoeve \emph{et~al.}, ``Revisiting the sequence constraint,'' in
  \emph{Proceedings of the 2006 International Conference on Principles and
  Practice of Constraint Programming}, Sep. 2006, pp. 620--634.

\bibitem{Apt2003}
K.~R. Apt, \emph{Principles of Constraint Programming}.\hskip 1em plus 0.5em
  minus 0.4em\relax Cambridge: Cambridge University Press, 2003.

\bibitem{GouldenJ1979}
I.~P. Goulden and D.~M. Jackson, ``An inversion theorem for cluster
  decompositions of sequences with distinguished subsequences,'' \emph{Journal
  of the London Mathematical Society}, vol.~2, no.~3, pp. 567--576, Dec. 1979.

\bibitem{Stanley2013}
R.~P. Stanley, \emph{Algebraic Combinatorics: Walks, Trees, Tableaux, and
  More}.\hskip 1em plus 0.5em minus 0.4em\relax New York: Springer-Verlag,
  2013.

\bibitem{MarcusRS2001}
\BIBentryALTinterwordspacing
B.~Marcus, R.~Roth, and P.~Siegel, ``Introduction to coding for constrained
  systems,'' Oct. 2001. [Online]. Available:
  \url{http://www.math.ubc.ca/\%7Emarcus/Handbook/}
\BIBentrySTDinterwordspacing

\bibitem{Kozyakin2009}
V.~Kozyakin, ``On accuracy of approximation of the spectral radius by the
  {G}elfand formula,'' \emph{Linear Algebra and its Applications}, vol. 431,
  no.~11, pp. 2134--2141, Nov. 2009.

\bibitem{WynerZ1973}
A.~D. Wyner and J.~Ziv, ``A theorem on the entropy of certain binary sequences
  and applications--{I},'' \emph{{IEEE} Transactions on Information Theory},
  vol. IT-19, no.~6, pp. 769--772, Nov. 1973.

\bibitem{ZehaviW1988}
E.~Zehavi and J.~K. Wolf, ``On runlength codes,'' \emph{{IEEE} Transactions on
  Information Theory}, vol.~34, no.~1, pp. 45--54, 1988.

\bibitem{ChayatS1989}
N.~Chayat and S.~Shamai, ``Extension of an entropy property for binary input
  memoryless symmetric channels,'' \emph{{IEEE} Transactions on Information
  Theory}, vol.~35, no.~5, pp. 1077--1079, Sept. 1989.

\bibitem{Blahut1972}
R.~E. Blahut, ``Computation of channel capacity and rate-distortion
  functions,'' \emph{{IEEE} Transactions on Information Theory}, vol. IT-18,
  no.~4, pp. 460--473, Jul. 1972.

\bibitem{Arimoto1972}
S.~Arimoto, ``An algorithm for computing the capacity of arbitrary discrete
  memoryless channels,'' \emph{{IEEE} Transactions on Information Theory}, vol.
  IT-18, no.~1, pp. 14--20, Jan. 1972.

\end{thebibliography}

\end{document}